\documentclass[]{article}
\usepackage{amsmath,amssymb,amsfonts}
\usepackage{graphicx}
\usepackage{textcomp}
\usepackage{arxiv}

\usepackage{caption}
\usepackage{csquotes}
\usepackage{tablefootnote}
\usepackage{booktabs}
\usepackage{glossaries-extra}
\usepackage{tabularx}
\usepackage{multirow}
\usepackage{bm}
\usepackage[backend=biber,backref=true]{biblatex}
\usepackage{varioref}
\usepackage[]{hyperref}
\usepackage{algorithm,algpseudocode}
\usepackage[]{cleveref}
\graphicspath{{./pics/}}

\crefname{figure}{Figure}{Figures}
\crefname{equation}{Equation}{Equations}
\crefname{table}{Table}{Tables}
\algrenewcommand\algorithmicindent{1.0em}%

\newcommand*\myemph[1]{``{#1}''}
\newcommand*\onedarray[1]{\bm{#1}}
\newcommand*\normal[1]{\overline{#1}}

\DeclareMathOperator*{\argmin}{arg\,min}
\newcommand{\lossfun}{\mathcal{L}(W_t, \widetilde{W}_t, \mathbf{h})}
\newcommand{\latspace}{\mathbf{h}}
\newcommand{\tablefootnotemark}[1]{\textsuperscript{\getrefnumber{#1}}}
\DeclareRobustCommand{\glsmes}[1]{\ensuremath{\text{\glsentryshort{#1}}}}

\DeclareRobustCommand{\orcidid}[1]{\href{#1}{\includegraphics[scale=.055]{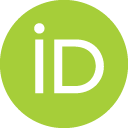}}}

\hypersetup{
pdftitle={Autoencoder based iterative modeling and multivariate time-series subsequence clustering algorithm},
pdfsubject={Arxiv preprint},
pdfauthor={J. Köhne, L. Henning, C. Gühmann},
pdfkeywords={Condition based maintenance, Multivariate time-series data, Change point detection, Unsupervised clustering, Autoencoder, Segmentation, Subsequence, Clustering},
}

\title{Autoencoder Based Iterative Modeling and Multivariate Time-Series Subsequence Clustering Algorithm}

\author{%
    \href{https://orcid.org/0000-0001-5536-9604}{\includegraphics[scale=.055]{pics/ORCIDiD_icon128x128.png}} Jonas Köhne \\
	Chair of Electronic Measurement and Diagnostic Technology\\
	Technische Universität Berlin\\
	Berlin, Germany\\
	\texttt{j.koehne@tu-berlin.de} \\
	\And
	\href{https://orcid.org/0000-0001-6724-9781}{\includegraphics[scale=.055]{pics/ORCIDiD_icon128x128.png}} Lars Henning \\
	Commercial Vehicle Electronics / Powertrain \& Power Engineering\\
	IAV GmbH\\
	Berlin, Germany\\
	\texttt{lars.henning@iav.de} \\\And
	\href{https://orcid.org/0000-0002-2865-0078}{\includegraphics[scale=.055]{pics/ORCIDiD_icon128x128.png}} Clemens Gühmann \\
	Chair of Electronic Measurement and Diagnostic Technology\\
	Technische Universität Berlin\\
	Berlin, Germany\\
	\texttt{clemens.guehmann@tu-berlin.de} \\
}

\renewcommand{\shorttitle}{ABIMCA}
\newglossaryentry{ms}{name=mechatronic system, description={A combination of mechanics electronics and information technology}}
\newglossaryentry{md}{name=machine diagnosis,description={}}
\newglossaryentry{ml}{name=machine learning,description={}}
\newglossaryentry{ts}{name=time-series,description={}}
\newglossaryentry{ch}{name=calinski-harabasz,description={}}
\newglossaryentry{db}{name=davies-bouldin,description={}}
\newglossaryentry{sil}{name=silhouette,description={}}
\newglossaryentry{kmeans}{name=\textit{k}-means,description={}}

\newabbreviation{bae}{BAE}{Base Autoencoder}
\newabbreviation{sae}{SAE}{Subsequence Autoencoder}
\newabbreviation{cbm}{CbM}{condition-based maintenance}
\newabbreviation{iid}{i.i.d}{independent and identically distributed}
\newabbreviation{mse}{MSE}{mean squared error}
\newabbreviation{mt3sid}{MT3SID}{multivariate time-series sub-sequence identification and discovery}
\newabbreviation{mt3scm}{\textit{MT3SCM}}{Multivariate Time-Series Sub-Sequence Clustering Metric}
\newabbreviation{abimca}{ABIMCA}{Autoencoder Based Iterative Modeling and Subsequence Clustering Algorithm}
\newabbreviation{ae}{AE}{Autoencoder}
\newabbreviation{ar}{AR}{auto regression}
\newabbreviation{hmm}{HMM}{Hidden Markov Model}
\newabbreviation{mcmc}{MCMC}{Markov chain Monte Carlo}
\newabbreviation{dtw}{DTW}{Dynamic Time Warping}
\newabbreviation{tsd}{TSD}{\gls{ts} data}
\newabbreviation{mts}{MTS}{multivariate \gls{ts}}
\newabbreviation{mtsd}{MTSD}{multivariate \glsxtrlong{tsd}}
\newabbreviation{sts}{STS}{short \gls{ts}}
\newabbreviation{mc}{MC}{Markov chain}
\newabbreviation{cp}{CP}{change-point}
\newabbreviation{bocpd}{BOCPD}{Bayesian Online \Glsxtrlong{cp} Detection}
\newabbreviation{cc}{$cc$}{curvature consistency}
\newabbreviation{wcc}{$cc_w$}{weighted curvature consistency}
\newabbreviation{sl}{$s_L$}{silhouette location based}
\newabbreviation{sp}{$s_P$}{silhouette curve-parameter based}
\newabbreviation{sgd}{SGD}{stochastic gradient descent}
\newabbreviation{RNN}{RNN}{recurrent neural network}
\newabbreviation{GRU}{GRU}{gated recurrent unit}
\newabbreviation{FNN}{FNN}{feedforward neural network}
\newabbreviation{CNN}{CNN}{convolutional neural network}
\newabbreviation{MLP}{MLP}{multilayer perceptron}
\newabbreviation{DBN}{DBN}{deep belief network}
\newabbreviation{GAN}{GAN}{generative adversarial network}
\newabbreviation{VAE}{VAE}{variational autoencoder}
\newabbreviation{RDE}{RDE}{Real Driving Emissions}
\newabbreviation{WLTC}{WLTC}{Worldwide harmonized Light Duty Test Cycle}
\newabbreviation{VAR}{VAR}{vector autoregression}
\newabbreviation{PCA}{PCA}{principal component analysis}
\newabbreviation{birch}{BIRCH}{Balanced Iterative Reducing and Clustering using Hierarchies}
\newabbreviation{clustream}{CluStream}{Stream Clustering Framework}
\newabbreviation{dbstream}{DBSTREAM}{Density-based Stream Clustering}
\newabbreviation{sostream}{SOStream}{Self Organizing density-based clustering over data Stream}
\newabbreviation{denstream}{DenStream}{Density-Based Clustering over an Evolving Data Stream with Noise}
\newabbreviation{cf}{CF}{clustering features}
\newabbreviation{dbscan}{DBSCAN}{Density-Based Spatial Clustering of Applications with Noise}
\newabbreviation{minibatchkmeans}{MiniBatchKMeans}{Mini-Batch K-Means}
\newabbreviation{mocap}{MOCAP}{The Motion Capture Database}

\begin{document}
\maketitle
\begin{abstract}
    This paper introduces an algorithm for the detection of change-points and the identification of the corresponding subsequences in transient \gls{mtsd}.
    The analysis of such data has become more and more important due to increasing availability in many industrial fields.
    Labeling, sorting or filtering highly transient measurement data for training \gls{cbm} models is cumbersome and error-prone.
    For some applications it can be sufficient to filter measurements by simple thresholds or finding change-points based on changes in mean value and variation.
    But a robust diagnosis of a component within a component group for example, which has a complex non-linear correlation between multiple sensor values, a simple approach would not be feasible.
    No meaningful and coherent measurement data which could be used for training a \gls{cbm} model would emerge.
    Therefore, we introduce an algorithm which uses a \gls{RNN} based \gls{ae} which is iteratively trained on incoming data.
    The scoring function uses the reconstruction error and latent space information.
    A model of the identified subsequence is saved and used for recognition of repeating subsequences as well as fast offline clustering.
    For evaluation, we propose a new similarity measure based on the curvature for a more intuitive \gls{ts} subsequence clustering metric.
    A comparison with seven other state-of-the-art algorithms and eight datasets shows the capability and the increased performance of our algorithm to cluster \gls{mtsd} online and offline in conjunction with mechatronic systems.
\end{abstract}
\glsresetall

\section{Motivation}\label{sec:Motivation}
In the applications of \gls{md} of \glsplural{ms} and the subfield \gls{cbm}, all supervised methods rely on high-quality labeled data \cite{Quatrini.2020,Fault.2006}.
But also, the unsupervised methods are more robust and provide better results when a \myemph{good} distribution of the data preexists.
Unsupervised \gls{ml} methods generally require \gls{iid} data to produce acceptable results.
Providing data with this distribution in an industrial measurement setup is impractical, especially when continuously changing conditions cause time dependency.
An option for a \gls{ms} with different operating points is to measure the main operating points separately and create a diagnosis method for each of those individually.
This requires a well-structured design and execution of experiments with a measurement labeling process.
In a real world development environment for \gls{ms}, where measurements are taken either automatically and/or manually and by many individuals and in different hardware and software development stages, providing consistently labeled and categorized data is a challenge.

Automatically labeling and categorizing \gls{mts} data is therefore not only an alleviation but might be crucial for a successful \gls{cbm} approach.
As described above, labeled and categorized data is essential for training a model to represent a \gls{ms} in a data driven approach.
In the automotive sector where a lot of measurements occur at different operating points this is especially important.
Some of these measurements are being recorded on a test bench in standard environment conditions with predefined operating points and for a given time (e.g., \gls{WLTC}).
Others can be in idle mode during waiting or preparation time for a longer trip or other measurements.
Also, very transient episodes are existent (e.g., \gls{RDE}).
All of these measurements do not necessarily have the same calibration of the underlying \gls{ms}.
To train a robust model of the \gls{ms}, component group or a single component, a big effort has to be put in the design of the experiments alone, not to mention the experiments themselves.
Therefore, a method of automatically labeling existing measurements is of advantage.
Afterwards an automatic sorting of the labeled time sequences by statistical methods is possible, to enable a data driven mechatronic diagnosis approach.

Using advanced unsupervised approaches for \gls{cbm} allows the data to be unlabeled (otherwise supervised methods could be used).
In this case the labeling refers to the label of the \emph{condition} (mechanical degradation) of the monitored system.
When trying to diagnose \glsplural{ms} that have many operating points and are free to transfer in between those or are capable of totally transient operation modes, then a robust diagnosis of the actual \emph{condition} of the \gls{ms} is extremely challenging.
An early and reliable (robust) diagnosis of a \gls{ms} prevents accidents, enables optimal maintenance and increases uptime of machinery.
Without the knowledge of the current \emph{condition} of the system, fault prevention can only be done by predetermined maintenance intervals.
Motivation is therefore to monitor the health \emph{condition} of the \gls{ms} as close as possible, resulting in the task of separating discrete sensory data into uniquely identifiable and recognizable segments or subsequences.
This is beneficial to the performance of \emph{anomaly detection}, because if all normally occurring subsequences are identified, the detection of abnormal or faulty subsequences is straightforward.

\section{Introduction}\label{sec:Introduction}
When monitoring the health condition of a \gls{ms} it is state of the art, to manually calibrate specific release conditions, during which the condition monitoring is enabled.
This is done to exclude operating points which are very rare, too transient or are just not feasible for drawing conclusions about the \emph{condition} of the \gls{ms}.
But even restricting the conditions on where to diagnose the machine (which already is reducing the probability of diagnosing the machine at all due to an operating state which is by chance outside the release conditions) cannot always help to improve the fault detection, identification and quantification of its magnitude.
For example, in a \gls{ms} with a complex nonlinear dependency of its subcomponents and its time dependency, \myemph{going in} or \myemph{going out} of the release conditions can result in very different system behavior.
Comparing these two states does not lead to reliable conclusions for the \glsplural{ms} health \emph{condition}.

Therefore, it is crucial for any monitoring strategy what kind of data sequences are used for training/calibration and validation.
Manually screening, labeling and sorting data into comparable sequences is time-consuming, error-prone and cumbersome.
Additionally, this is a decision process which requires expert and domain knowledge.

The new algorithm which we introduce in this work is capable of generating subsequence models from online streaming data which is processed sequentially.
Any coherent subsequence that is identified can be recognized (clustered) if occurring again.
Depending on multiple \emph{sensitivity} calibration parameters, time-varying data points are associated and identified as a subsequence.
The parameters determine the \emph{volatility} or the strength of the affiliation required to be recognized as one time varying subsequence.
These subsequence models can also be applied efficiently offline onto large existing datasets.
During this prediction phase, the algorithm provides a vector of subsequence labels which were recognized as one from the training data.
Depending on the calibration, it can also provide a label for \emph{unknown} data which represents a phase where no pattern could be recognized.
Otherwise, it finds the best fitting subsequence and labels it as that.
The approach published in this work is currently only based on \gls{mts} input but could be adapted for a univariate input.
It is a multivariate time-series sub-sequence discovery and identification method.

Our contribution is a new algorithm for online subsequence clustering of \gls{mtsd} called \myemph{\gls{abimca}} and a new metric to evaluate cluster algorithms focused on this task \myemph{\gls{mt3scm}}.
We compare our algorithm with
\begin{itemize}
    \item seven other state-of-the-art algorithms
    \item eight datasets from which six are publicly available and two are provided with our codebase
    \item three widely used unsupervised clustering metrics
    \item our own metric (\gls{mt3scm}) and its four components
\end{itemize}
while varying the use of default algorithm parameters with optimized parameters on each algorithm and dataset via random grid search.


\section{Related work}\label{sec:Related-work}
In this section we define the terminology used and its semantics to categorize our work within the large bibliography existing in this field and provide a selected list of related works and their ascendancy to this paper.

\subsection{Terminology and Semantics}\label{subsec:Terminology-and-Semantics}
Numerous possibilities have been described for achieving our main goal of segmenting discrete \gls{ts} sensory data.
Most approaches can be sorted into the following partially overlapping categories: \emph{\gls{ts} analysis}~\cite{Time.2016}, \emph{pattern recognition}~\cite{Bishop.2016}, \emph{temporal knowledge discovery}~\cite{Roddick.2002}, \emph{motif discovery}~\cite{Lin.2002}, \emph{\glsxtrlong{cp} detection}~\cite{vandenBurg.2020}, \emph{data clustering}~\cite{Jain.1999} or \emph{anomaly detection}~\cite{Chandola.2009}.
All those terms refer to methods or algorithms which could be used directly or indirectly to achieve our goal.
Explicit description on each term or category can be found by the interested reader in the stated references.

To limit the scope of this work, we focus on \emph{data clustering} which can be separated into six subcategories by the following groups of two: \emph{univariate} and \emph{multivariate data}, \emph{online} and \emph{offline} algorithms, \emph{variant} and \emph{invariant} data.
\begin{figure}[h]
    \centering
    \def\svgwidth{.5\columnwidth}
    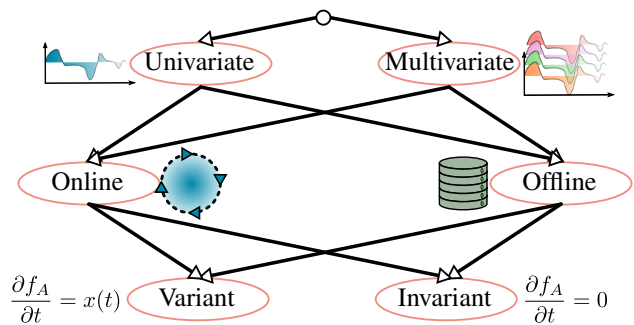
    \caption{Combination possibilities of \gls{ts} clustering categories.}
    \label{fig:time-series-categories}
\end{figure}
The term \myemph{clustering} implies an unsupervised method. The equivalent supervised method would be called \myemph{classification}.
In the relevant literature more and other distinctions are made, depending on the specific field and context.
First, we describe \gls{tsd} since this is the data format we focus on in this paper.
Afterwards we explain the differences between the subcategories.

``A \gls{ts} is a sequence of observations taken sequentially in time.''~\cite*{Time.2016}
We denote a \gls{ts} data point as an observation with exactly one connected timestamp.
The timestamp is not a variable or feature.

\paragraph{univariate -- multivariate}
If a single value or a scalar is the only variable of the data, then the data is univariate.
It is the most basic format data can have.
Considering \gls{tsd}, a single temperature sensor with a timestamp would be univariate.
Univariate \gls{tsd} could also be interpreted as multivariate data of two dimensions, when taking the timestamp as another variable or feature.

\paragraph{online -- offline}
The differentiation between online and offline algorithms or analysis of data is crucial.
Offline refers to data analysis that is applied to all the data at once.
Measurement data, for example, is available in one or multiple files or can be accessed via a previously filled database.
Offline algorithms can therefore iterate and optimize their result based on a criterion applied to known data.
Online analysis on the other hand, is applied sequentially.
The algorithm needs to be able to function with a criterion that generalizes well on unknown data.
One chunk or piece of data can be applied on the online algorithm without knowing the rest of the data.
This approach cannot be as robust and accurate as an offline analysis, which is why most methods found in the literature are offline algorithms.
Ideally the online algorithm learns every time new data is provided.
For the sake of completeness, however, it should be noted that some online algorithms have to be pretrained offline and some algorithms referred to as offline, can be used sequentially.
This depends on the underlying methods used.
An offline algorithms' purpose is not to be used online.
Online algorithms, nevertheless, can be used offline.

\paragraph{dependent -- independent \gls{tsd} (time variant -- time invariant)}
Depending on the field of study the specific terminology of time dependency can differ.
We want to emphasize on the common accepted assumption \blockquote[{\cite[S.~1]{Time.2016}}]{An intrinsic feature of a \gls{ts} is that, typically, adjacent observations are dependent}.
Time dependency characterizes \gls{tsd}, where a consecutive observation has some connection with its predecessor.
In some fields a connection is not necessarily given for two data points in a database that are the closest to each other regarding their timestamp.
The dependency of adjacent observations is self-evident, when collecting sensor values of a mechatronic system from an experimental rig, for example.
We therefore use the term \myemph{time dependency} in the context of a dynamical system and extend it to a time-variant system in the terminology of control systems engineering.
Independent \gls{tsd} would be where the variance and the average are invariant along the time (stationary) and ergodic.

We position our work in the subcategory of \textbf{online clustering of dependent \glsxtrlong{mtsd}}.
As of now we refer to a \gls{ts} as a sequence of dependent observations with a constant sample-rate.
A \myemph{discrete series of consecutive data points} as a subset of this \gls{ts} is synonymously referred to as \emph{pattern, motif, sequence, operating state, state change, between change points, subsequence, episode} or \emph{segment}, among others.
In this paper we will use the term \emph{\textbf{subsequence}} (using terminology of \cite{Agrawal.1995}).

\subsection{Algorithms and datasets}\label{subsec:AlgorithmsAndDatasets}
In the relevant literature a diverse number of clustering algorithms can be found~\cite{Xu.2015}.
Due to this fact, most of the existing literature for reviewing or surveying existing approaches, attend to a higher-level scope~\cite{Lovric.2014, Torkamani.2017, Aminikhanghahi.2017}.
\begin{table}[h!]
    \small
    \centering
    \caption{Algorithms used for \gls{ts} clustering comparison}
    \label{tab:algorithms}
    \begin{tabular}{llll}
    \toprule
        & type & library & publication \\
    algorithm &  &  &  \\
    \midrule
    BIRCH\tablefootnote{\url{https://scikit-learn.org/stable/modules/classes.html}\label{footnote:sklearn}} & hierarchical & sklearn~\cite{Pedregosa.2011} & \cite{Zhang.1996} \\
    BOCPD\tablefootnote{\url{https://facebookresearch.github.io/Kats/}} & distribution-based & kats~ & \cite{Adams.19.10.2007} \\
    CluStream\tablefootnote{\url{https://riverml.xyz}\label{footnote:riverml}} & density-based & river~\cite{Montiel.08.12.2020} & \cite{Aggarwal.2003} \\
    DBSTREAM\tablefootnotemark{footnote:riverml} & density-based & river~\cite{Montiel.08.12.2020} & \cite{Hahsler.2016} \\
    DenStream\tablefootnotemark{footnote:riverml} & density-based & river~\cite{Montiel.08.12.2020} & \cite{Cao.2006} \\
    MiniBatchKMeans\tablefootnotemark{footnote:sklearn} & distance-based & sklearn~\cite{Pedregosa.2011} & \cite{Sculley.2010} \\
    STREAMKMeans\tablefootnotemark{footnote:riverml} & distance-based & river~\cite{Montiel.08.12.2020} & \cite{Montiel.08.12.2020} \\
    \bottomrule
    \end{tabular}
\end{table}
Fewer are concentrating on \gls{ts} clustering~\cite{Liao.2005, Zolhavarieh.2014,Aghabozorgi.2015}, online~\cite{Carnein.2019}, temporal knowledge discovery~\cite{Roddick.2002}, sequential pattern recognition~\cite{Agrawal.1995}, high dimensional data~\cite{Kriegel.2009} or change point detection~\cite{Aminikhanghahi.2017,Truong.2020}.
Current approaches use deep learning architectures like \gls{MLP}, \gls{CNN}, \gls{DBN}, \gls{GAN} and \gls{VAE} among others~\cite{Aljalbout.2018}.

The algorithms we use for comparison are listed in~\cref{tab:algorithms}.
All are online clustering algorithms, that can be used for \gls{ts} clustering, implementations are publicly available in the Python programming language (see \emph{library} column in~\cref{tab:algorithms}) and are well established and tested.

The \gls{birch}~\cite{Zhang.1996} algorithm is based on a \gls{cf} tree with the \gls{cf} as a triple of the number of data points, linear sum and the squared sum.
This CF tree is built dynamically.
It was also one of the earliest algorithms capable of online clustering.
The \gls{bocpd}~\cite{Adams.19.10.2007} algorithm as the name suggests, uses Bayesian methods to detect \glspl{cp} online.
Since this algorithm only detects \glspl{cp}, we manipulated the result to be able to interpret every \gls{cp} as the beginning of a new cluster.
This algorithm starts in our comparison with the limitation of not being able to recognize a previously seen cluster.
The \gls{clustream}~\cite{Aggarwal.2003} algorithm is based on extended \gls{cf} from \gls{birch}, following a \gls{kmeans} algorithm.
The \gls{dbstream}~\cite{Hahsler.2016} algorithm is based on the \gls{sostream}~\cite{Isaksson.2012} and uses a shared density graph to capture the density between micro-clusters.
The \gls{denstream}~\cite{Cao.2006} algorithm  is an extension of \gls{dbscan}~\cite{Ester.1996} which uses a damped window model of \gls{cf} to create core-micro-clusters and outlier-micro-clusters.
The \gls{minibatchkmeans}~\cite{Sculley.2010} algorithm proposes \blockquote[{\cite{Sculley.2010}}][]{the use of mini-batch
optimization for k-means clustering} to improve the \gls{kmeans} optimization problem.
The STREAMKMeans~\cite{Montiel.08.12.2020} algorithm uses an adaption of the original STREAM algorithm from~\cite{OCallaghan.2002}. Replacing the \textit{k}-median subroutine LSEARCH by an incremental \gls{kmeans} algorithm.
More information and comparison of most of the used algorithms can be found in \cite{Carnein.2019, Ghesmoune.2016}

As described in \cref{sec:Introduction}, the focus of this publication is on the online multivariate time dependent subsequence clustering using \gls{RNN} based \gls{ae}.
The number of algorithms within this scope is limited compared to the number of clustering algorithms in general.
The following approaches use at least some of those prerequisites.
\cite{Keogh.2004b} emphasize the term \emph{segmentation} for an offline sliding window and bottom-up algorithm.
Others are converting the \gls{ts} into a \gls{mc} and then using a Bayesian method to cluster the \glsplural{mc}~\cite{Ramoni.2002}.
Referring them as \emph{episodes}.
Here the data needs to be discretized into bins of equal length.
Reference \cite{Wang.2006} use manually selected characteristics (e.g., kurtosis, skewness and frequency) for clustering univariate \gls{tsd}.
Others are using the Augmented Dickey-Fuller test to evaluate \gls{ts} stationarity and perform a segmentation based on this~\cite{Silva.2021}.
In \cite{Lu.2020} dynamic latent variables from a \gls{VAR} model in combination with a \gls{PCA} is used for segmenting industrial \gls{tsd}.
Reference \cite{Ceci.2020} show the advantage of an embedding approach as well, by introducing a PCA and a Vanilla-\gls{ae} \gls{cp} detection method with the restriction of focusing on multivariate power grid data.

Focused on transfer learning, \cite{Li.2022} introduce an adversarial approach for domain adaption using a stacked \gls{ae}.
Offline convolutional sparse \gls{ae} used for supervised sequence classification was done by~\cite{Baccouche.2012} and adapted by~\cite{Bascol.2016} for unsupervised motif mining.
Other \gls{ae} based papers are, for example, by \cite{Chazan.2019} who use a mixture of \glspl{ae} for image and text clustering.
Stacked \gls{ae} and \gls{kmeans} for offline clustering is done by \cite{BoYang.2017} without considering time dependency.
Showing the combination of GRU-based \gls{ae} and \gls{mts} for anomaly detection is done by \cite{Guo.2018}.
Reference \cite{Chen.2020} apply the sliding windows approach on \gls{CNN}-based \gls{ae} for Anomaly Detection of Industrial Robots.
A similar approach using \gls{ae} for \gls{mts} segmentation is published in~\cite{Lee.2018}. The focus there is on change point detection using latent space variables only and no clustering or identification of the subsequences is done.

Clustering is strongly depending on the data and task provided, as \citeauthor*{Jain.1999} expressed in \citeyear{Jain.1999}:
\blockquote[{\cite[S.~268]{Jain.1999}}][.]{\dots; each new clustering algorithm performs slightly better than the existing ones on a specific distribution of patterns}
Therefore, we try to apply the algorithms on multiple different \gls{mts} datasets and compute different metrics for comparison.
Large efforts are made for making datasets available to the scientific community and the public to improve comparability and reproducibility by universities or governmental institutions~\cite{Dua.2017,Data.gov.14.09.2020}.
For this paper we focus on data with multivariate quantitative features with continuous values. For a list of the datasets see~\cref{tab:datasets} and a brief description is given in \vref{sec:Datasets}.
Evaluating the performance of a clustering algorithm can be done with two different approaches.
If external knowledge about the ground truth of each data point and its cluster is known, then so-called external measures can be applied.
If no ground truth is available, internal measures need to suffice.
Many external measures exist, like the well-known F1-score (based on the effectiveness measure by~\cite{vanRijsbergen.2008}).
With the large number of data available and working in the context of transient machine behavior with the focus on finding internal states of the system, acquiring or providing the ground truth is time-consuming, error-prone and cumbersome (as described in \cref{sec:Introduction}).
\blockquote[{\cite[S.~30]{Aghabozorgi.2015}}][.]{The definition of clusters depends on the user, the domain, and it is subjective}.
We therefore use internal measures for comparing our approach.
Those internal measures commonly rely on a similarity measure of the actual data which being clustered.
Using the provided label from the cluster algorithm many possible combinations of, for example, the density of all points in a cluster and the distance to the center of the nearest other cluster are available.
Thorough work on metric comparison and similarity measures has been done~\cite{Kremer.2011, Serra.2014, Aghabozorgi.2015}.
Most of those measures are based on simple distances and densities computed for each data point but do not take time dependency into consideration.
Because of this, we found, that for the use case described in this paper, the commonly used clustering evaluation measures are not well suited for \myemph{\gls{ts} clustering evaluation measures}.
In \cref{sec:Metric} we introduce an approach for similarity measures which considers time dependency in combination with well-established clustering metrics (see~\cref{tab:metrics}).
\begin{table}[h!]
    \small
    \centering
    \caption[Metrics used for \gls{ts} clustering comparison]{Metrics used for \gls{ts} clustering comparison. Implementations used from~\cite{Pedregosa.2011}}
    \label{tab:metrics}
    \begin{tabularx}{\columnwidth}{>{\hsize=.5\hsize}XXX}
    \toprule
                        & valuation & value \\
    metric              &  &  \\
    \midrule
    \gls{sil}~\cite{Rousseeuw.1987}           & Ratio of distance to its own cluster and distance to the nearest cluster center& $\left\{s\in \mathbb{R} : -1 \leq s \leq 1 \right\}$ the higher, the better\\
    \gls{ch}~\cite{Calinski.1974}   & Ratio of between-cluster variance and the within-cluster variance& $\left\{s\in \mathbb{R} : 0 \leq s \leq \infty \right\}$ the higher, the better \\
    \gls{db}~\cite{Davies.1979}    & Ratio of cluster size and between-cluster distance& $\left\{s\in \mathbb{R} : 0 \leq s \leq \infty \right\}$ the lower, the better\\
    \bottomrule
    \end{tabularx}
\end{table}
\begin{table}[h!]
    \small
    \centering
    \caption{Datasets used for \gls{ts} clustering comparison}
    \label{tab:datasets}
    \begin{tabular}{llll}
        \toprule
            & type & features & publication \\
        dataset &  &  &  \\
        \midrule
        bee-waggle\tablefootnote{\url{https://sites.cc.gatech.edu/~borg/ijcv\_psslds/}} & feature extraction from video & 4 & \cite{Oh.2008} \\
        cmapss\tablefootnote{\url{https://data.nasa.gov/dataset/C-MAPSS-Aircraft-Engine-Simulator-Data/xaut-bemq}} & simulation & 18 & \cite{AriasChao.2021} \\
        eigen-worms\tablefootnote{\url{http://www.timeseriesclassification.com/description.php?Dataset=EigenWorms}} & feature extraction from video & 6 & \cite{Yemini.2013} \\
        hydraulic\tablefootnote{\url{https://archive.ics.uci.edu/ml/datasets/Condition\%20monitoring\%20of\%20hydraulic\%20systems}} & test rig sensors & 17 & \cite{Schneider.2017} \\
        lorenz-attractor & computation & 3 & \cite{Lorenz.1963} \\
        mocap\tablefootnote{\url{http://mocap.cs.cmu.edu/}} & motion capturing sensors & 93 & \cite{mocap.21.04.2022} \\
        occupancy\tablefootnote{\url{https://github.com/LuisM78/Occupancy-detection-data}} & measurement & 5 & \cite{Candanedo.2016} \\
        thomas-attractor & computation & 3 & \cite{THOMAS.1999} \\
        \bottomrule
    \end{tabular}
\end{table}

\section{Definitions and Restrictions}\label{sec:Definitions-and-Restrictions}
In the following section we define in more detail our data, together with the restrictions of our environment.
Considering online clustering, we can refer to our \gls{tsd} as continuously incoming data or streaming data.
This data is considered multivariate when the dimension (number of sensors or features) of the data stream $d > 1$.
When we denote one value of one feature as $x$, we have at time step $t$ the following feature vector:
\begin{align}
	\onedarray{x}_{t} &  = \left(x_0, x_1, \ldots, x_{d}  \right)^T \quad \text{with } x \in \mathbb{R} \text{ and } d \in \mathbb{N}\label{eq:colvector}
\end{align}
Whereas the natural numbers include zero $ \left\{0, 1, 2, \ldots\right\} = \mathbb{N}$.
A complete measurement sequence with $n$ number of time steps using $\onedarray{x}$ from \cref{eq:colvector} as
\begin{align}
	X &  = \left( \onedarray{x}_0, \onedarray{x}_1, \ldots, \onedarray{x}_{n} \right) \quad \text{with } n \in \mathbb{N} \label{eq:sequencedata}
\end{align}
so, $X \in \mathbb{R}^{d \times n}$.
With time dependency consideration, it is reasonable to denote a sliding window of the streaming data, considering \cref{eq:colvector} and $n$ the number of samples already collected as:
\begin{align}
	W_t &  = \left(\onedarray{x}_{t+0}, \onedarray{x}_{t+1}, \ldots, \onedarray{x}_{t+\zeta} \right)  \Rightarrow  (\zeta \in \mathbb{N}) \land (\zeta < n) \label{eq:slidingwindow}
\end{align}

Let's also assume, that within the measurement data $X$ there exist subsequences $S_j$ which satisfy our requirements of non-overlapping and variable length.
For the indexes of our subsequences, we denote
\begin{align}
	\mathcal{J} & = \left\{j\in \mathbb{N} : j \leq u \right\} \label{eq:subsequenceindex}
\end{align}
where $u$ is the number of identified subsequences in $X$.
The subsequence then is a continuous sampling from $X$ for a small period of time steps with consecutive data points and the length of $m$.
The subsequence length is usually much smaller than the length of the full measurement data $m \ll  n$.
\begin{align}
	S_{j} &  = \left(\onedarray{x}_{q_{j}}, \ldots, \onedarray{x}_{q_{j} + m_{j}}  \right)  \quad (0 \leq q \leq n-m) \land j \in \mathcal{J} \label{eq:subsequence1data}
\end{align}
For each subsequence with the index $j$ we have a first time step index $q_{j} = q_{j,start}$ and a last $q_{j} + m  = q_{j,end}$ for which we do not allow overlapping
\begin{align}
        \begin{split}
	\forall j \in \mathcal{J} \quad \exists (q_{j, start}, q_{j, end})  \Rightarrow  \\ (q_{j, start} < q_{j, end}) \land((q_{j-1, end} < q_{j, start})  \land j > 0) \label{eq:asfasdf}
        \end{split}
\end{align}
This results in our uniquely identified non-overlapping set of subsequences
\begin{align}
	\mathcal{S} &  = \left\{S_0, \ldots, S_j  \right\} \quad j \in \mathcal{J}\label{eq:sequence}
\end{align}
Clustering these uniquely identified subsequences results in recognizing reoccurring subsequences and combining them into a subset of all subsequences
\begin{align}
	C_i & \subseteq \mathcal{S}\label{eq:cluster}
\end{align}
which results in the following cluster set $\mathcal{C}$
\begin{align}
	\mathcal{C} & = \left\{C_0, \ldots, C_i  \right\} \quad i \in \mathcal{I}
\end{align}
with the cluster index or unique cluster label
\begin{align}
	\mathcal{I} & = \left\{i\in \mathbb{N} : i < n \right\} \label{eq:clusterindex}
\end{align}

For the output of a clustering algorithm at time $t$ we denote the scalar value $y_t$ as our label or designated subsequence identification.
For evaluation purposes a clustering for a \gls{ts} produces a label array $\onedarray{y}$ for all time steps:
\begin{align}
	\onedarray{y} &  = \left(y_0, y_1, \ldots, y_{n}  \right) \quad \text{with } y \in \mathcal{J} \text{ and } n \in \mathbb{N}\label{eq:label_array}
\end{align}
Furthermore, it is a requirement, that the streaming data provided can be applied to a numerical differentiation algorithm. Therefore, a constant sample rate is necessary and in case of strong noise, filtering or smoothing of the data should be applied by a preprocessing step.
Also, there mustn't be missing values and extreme outliers need to be removed. In our use case we assume that some knowledge about the incoming data exists, so that an estimate of the variance and the mean of the variable can be performed for standardization.

\section{Datasets}\label{sec:Datasets}
All datasets used for comparison in this work are described briefly in this section and listed in~\cref{tab:datasets}.
They all contain quantitative features with continuous values.
For further use of the datasets, no missing values exist, the data is continuous and was standardized for the algorithms but not for the metric computations.
No other preprocessing like smoothening or filtering was performed.

The \textbf{bee-waggle} dataset~\cite{Oh.2008} contains movement of bees in a hive captured with a vision-based tracker.
The first two features are the x and y coordinates of the bee added with the sine and the cosine function applied to the heading angle.

The \textbf{cmapss} dataset is a ``dataset of run-to-failure trajectories for a small fleet of aircraft engines under realistic flight conditions''~\cite{AriasChao.2021} with 18 features.

The \textbf{eigen-worms} dataset~\cite{Yemini.2013} contains measurements of worm motion.
Preprocessing extracted six features, which represent the amplitudes along six previously identified base shapes of the worms

The \textbf{hydraulic} dataset~\cite{Schneider.2017} is obtained from a hydraulic test rig with measuring 17 process values such as pressures, volume flows and temperatures.

\begin{figure}[!t]
    \centering
    \includegraphics[width=.5\columnwidth]{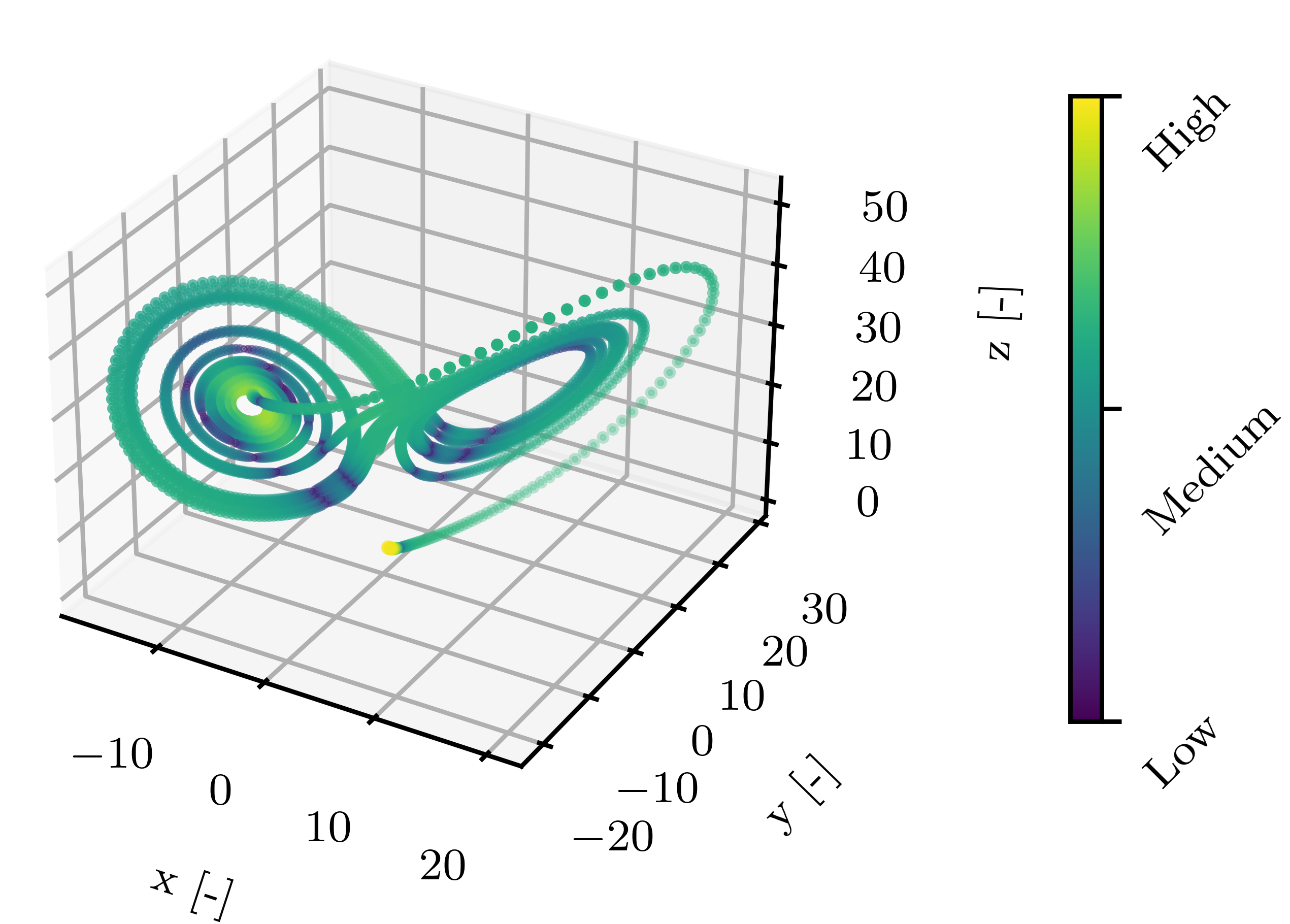}
    \caption{Lorenz-attractor dataset. Computed with $\dot{X} = s(Y - X); \dot{Y} = r X - Y - XZ; \dot{Z} = XY - b Z$ and parameters used $s=10$, $r=28$ and $b=2.667$. Color and marker size indicate amount of curvature on a logarithmic scale for better visibility.}
    \label{fig:lorenz-attractor-original}
\end{figure}

\textbf{Lorenz-attractor} refers to a synthetic dataset which is calculated using a system of the three coupled ordinary differential equations which represent a hydrodynamic system: $\dot{X} = s(Y - X); \dot{Y} = r X - Y - XZ; \dot{Z} = XY - b Z$ with parameters used $s=10$, $r=28$ and $b=2.667$ (see~\cref{fig:lorenz-attractor-original}).
``In these equations $X$ is proportional to the intensity of the convective motion, while Y is proportional to the temperature difference between the ascending and descending currents, similar signs of $X$ and $Y$ denoting that warm fluid is rising, and cold fluid is descending.''~\cite{Lorenz.1963}

The \textbf{mocap} or \gls{mocap} dataset~\cite{mocap.21.04.2022} contains 93 features from human motion captured with markers.

The \textbf{occupancy} dataset~\cite{Candanedo.2016} is a measurement of sensory data in an office with the following sensors: temperature, humidity, the derived humidity ratio, light and CO2.

The \textbf{thomas-attractor} dataset is as the lorenz-attractor dataset, a synthetic dataset, computed with the three coupled differential equations: $\dot{X} = \sin(Y) - bX; \dot{Y} = \sin(Z) - bY; \dot{Z} = \sin(X) - bZ$ originally proposed by~\cite{THOMAS.1999} with the parameter used $b=0.1615$.

\section{Multivariate Time-Series Sub-Sequence Clustering Metric (MT3SCM)}\label{sec:Metric}
As emphasized in \cref{sec:Introduction} and \cref{sec:Related-work}, to our knowledge none of the existing clustering metric exists, that takes the time space variations like curvature, acceleration or torsion in a multidimensional space into consideration.
We believe using these curve parameters, is an intuitive method to measure similarities between \gls{ms} state changes or subsequences in \gls{mtsd} in general (in regard to the restrictions in \cref{sec:Definitions-and-Restrictions}).

Our \gls{mt3scm} score consists of three main components.
\begin{align}
    mt3scm &= (\glsmes{wcc} + \glsmes{sl} + \glsmes{sp}) / 3\label{eq:mt3scm}
\end{align}
The \gls{wcc}, the \gls{sl} and the \gls{sp}.
When making the attempt of clustering \gls{tsd}, it is subjective and domain specific.
Nevertheless, we try to take the intuitive approach of treating \gls{mtsd} as space curves and use the parameterization as a similarity measure.
This is done in two different ways. 
First, we create new features by computing the curve parameters sample by sample (e.g., curvature, torsion, acceleration) and determine their standard deviation for each cluster.
Our hypothesis is, that with a low standard deviation of the curve parameters inside a cluster, the actions of a \gls{ms} in this cluster are similar.
We call this the \glsreset{cc}\gls{cc} (see \cref{eq:wcc} used in \cref{algo:ClusterCurvatureConsistency} in~\cref{algo:mt3scm}).
The second procedure is to apply these newly computed features, which are computed to scalar values per subsequence, onto a well-established internal clustering metric, the \gls{sil} score~\cite{Rousseeuw.1987} (see~\cref{tab:metrics}).

\begin{figure}[h!]
    \centering
    \includegraphics[width=.5\textwidth]{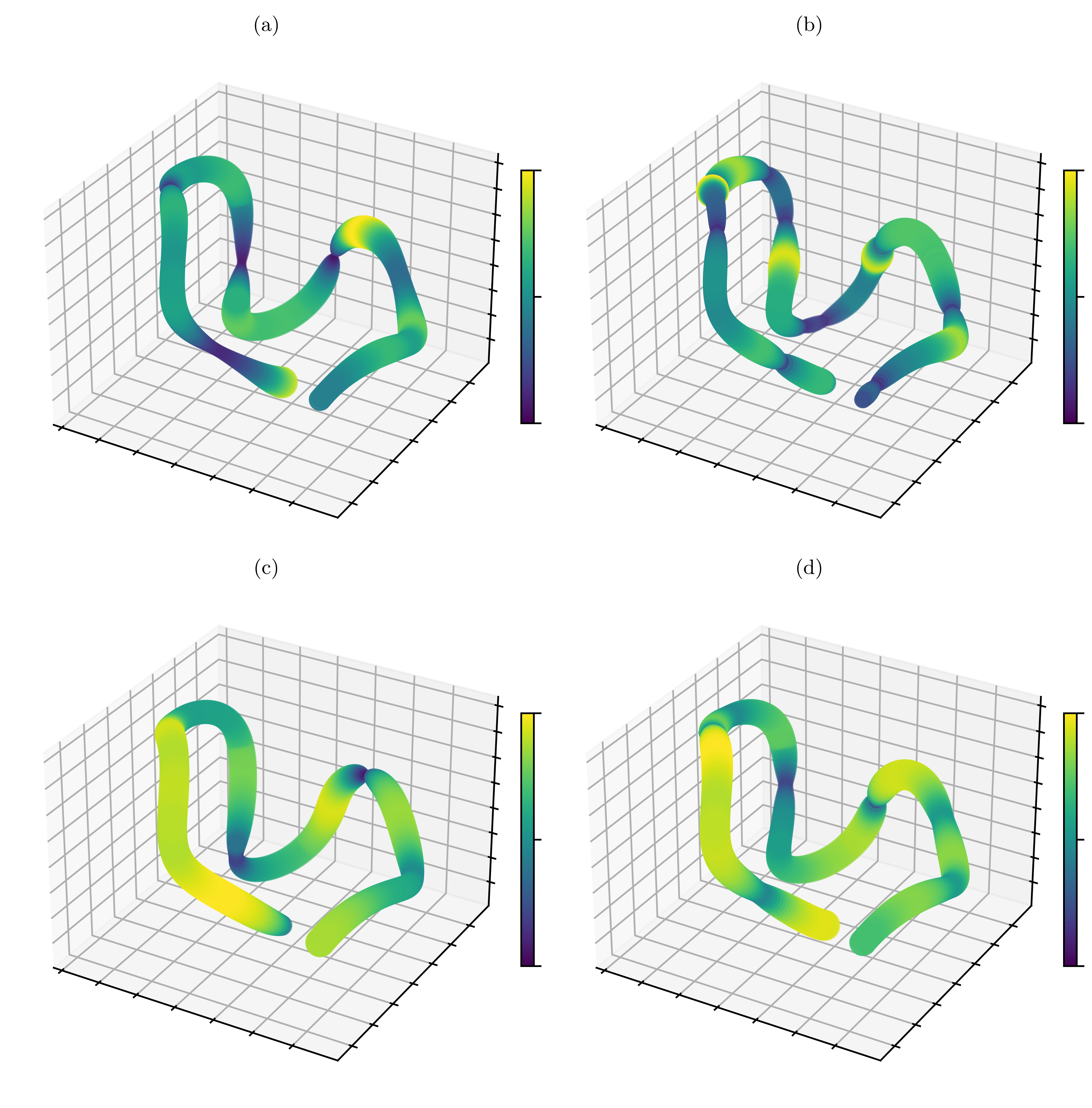}
    \caption[Curvature and Torsion example for metric calculation]{Qualitative visualization of the (a) curvature $\kappa$, (b) torsion $\tau$, (c) speed $v$ and (d) acceleration $a$ computed on part of the thomas-attractor dataset. Color and marker size indicate amount of curve parameter on a logarithmic scale for better visibility (dark and thin means low value, bright and thick means high value). Axis labels and colorbar labels are along the lines of \cref{fig:lorenz-attractor-original}.}
        \label{fig:curve}
\end{figure}

The computation of the \gls{cc} comprises the calculation of the curvature $\kappa$ and the torsion $\tau$ at every time step $t$ with~$\onedarray{x}_{t}$.
\begin{align}
    \kappa(t) & = \frac{\langle \dot{\onedarray{e}}_1(t), \onedarray{e}_2(t) \rangle}{\|\dot{\onedarray{{x}}}_{t}\|}
\end{align}
\begin{align}
    \tau(t) & = \frac{\langle \dot{\onedarray{e}}_2(t), \onedarray{e}_3(t) \rangle}{\|\dot{\onedarray{{x}}}_{t}\|}
\end{align}
Whereas $\onedarray{e}_1$ is the unit tangent vector (or first Frenet vector), $\onedarray{e}_2$ is the unit normal vector (or second Frenet vector) and $\onedarray{e}_3$ is the unit binormal vector (or third Frenet vector) which are defined as:
\begin{align}
    \onedarray{e}_1(t) & = \frac{\dot{\onedarray{{x}}}_{t}}{\|\dot{\onedarray{{x}}}_{t}\|}\\
    \onedarray{\normal{e}}_2(t) & = \ddot{\onedarray{{x}}}_{t} - \langle \ddot{\onedarray{{x}}}_{t}, \onedarray{e}_1(t) \rangle \times \onedarray{e}_1(t)\\
    \onedarray{e}_2(t) & = \frac{\onedarray{\normal{e}}_2(t)}{\|\onedarray{\normal{e}}_2(t)\|}\\
    \onedarray{\normal{e}}_3(t) & = \onedarray{\skew{-3}\dddot{{x}}}_{t} - \langle \onedarray{\skew{-3}\dddot{{x}}}_{t}, \onedarray{e}_1(t) \rangle \times \onedarray{e}_1(t) - \langle \onedarray{\skew{-3}\dddot{{x}}}_{t}, \onedarray{e}_2(t) \rangle \times \onedarray{e}_2(t)\\
    \onedarray{e}_3(t) & = \frac{\onedarray{\normal{e}}_3(t)}{\|\onedarray{\normal{e}}_3(t)\|}
\end{align}
From which we can also derive the speed $v = \|\dot{\onedarray{{x}}}_{t}\|$ and the acceleration $a = \|\ddot{\onedarray{{x}}}_{t}\|$.
\Cref{fig:curve} shows exemplarily the curvature $\kappa$, torsion $\tau$, speed $v$ and acceleration $a$ for the first part of the thomas-attractor dataset.

Afterwards the \gls{cc} is calculated per cluster $i \in \mathcal{I}$, by taking the empirical standard deviation for each curve parameter (exemplarily for $\kappa$ in \cref{eq:stdkappa} with the set of subsequence indexes $\mathcal{J}_i$ within our cluster $i$).
The arithmetic mean (\cref{eq:stdkappamean}) of the standard deviations for the curvature $\kappa$, torsion $\tau$ and the acceleration $a$ results in the final \gls{cc} per cluster (see \cref{eq:sccc}).
\begin{align}
    \sigma_{\kappa_i} &= \sqrt{\frac{1}{N_{i}-1}\sum_{j \in \mathcal{J}_i}  \sum _{n=q_{j}}^{q_{j}+m_{j}} \left(\kappa_{n}-{\overline{\kappa}_{i}}\right)^{2}}\label{eq:stdkappa}\\
    \sigma_{i} &= \frac{\sigma_{\kappa_{i}} + \sigma_{\tau_{i}} + \sigma_{a_{i}}}{3}\label{eq:stdkappamean}\\
    \glsmes{cc}_{i} &= 1 - \sigma_{i} \quad \text{with } \glsmes{cc}_{i} \in\mathbb{R}: \glsmes{cc}_{i}  \leq 1\label{eq:sccc}\\
    \glsmes{cc}_{i} &=
    \begin{cases}
        \glsmes{cc}_{i}, & \mbox{if } \glsmes{cc}_{i} > -1 \\
        -1,  & \mbox{if } \glsmes{cc}_{i} \leq -1 \\
    \end{cases}\label{eq:wcclimited}
\end{align}
The \gls{wcc}, is directly derived from the \gls{cc} per cluster, by weighting it with the number of data points per cluster $i \in \mathcal{I}$ \cref{eq:wcc}.
\begin{align}
    \glsmes{wcc} &= \frac{\sum\limits_{i=1}^{n} \glsmes{cc}_i \times N_{i}}{\sum\limits_{i=1}^{n} N_{i}} \label{eq:wcc}
\end{align}
The calculation of the scores \gls{sp} and \gls{sl} is different to the standard estimation of the  \gls{sil} score, which is shown in \cref{eq:standardsil} and originally based on every data point of the \gls{ts} $X$ and the assigned cluster label array $y$:
\begin{align}
    s &= f(X, \onedarray{y}) \label{eq:standardsil}
\end{align}
Our \gls{sp} is the \gls{sil} score derived from our previously computed curve parameters per subsequence per cluster as well as the standard deviation of those and the number of data points per subsequence.
\begin{align}
    \glsmes{sp} &= f(\widetilde{X}_{\glsmes{sp}}, \onedarray{y}_j) \label{eq:silparams}
\end{align}
with
\begin{align}
    \widetilde{X}_{\glsmes{sp}} = {
        \begin{pmatrix}
            \overline{\kappa}_{11} & \overline{\tau}_{11} & \overline{a}_{11} & \sigma_{11} & N_{11} \\
            \overline{\kappa}_{12} & \overline{\tau}_{12} & \overline{a}_{12} & \sigma_{12} & N_{12} \\
            \vdots                &\vdots               &\vdots            &\vdots      &\vdots \\
            \overline{\kappa}_{21} & \overline{\tau}_{21} & \overline{a}_{21} & \sigma_{21} & N_{21} \\
            \overline{\kappa}_{22} & \overline{\tau}_{22} & \overline{a}_{22} & \sigma_{22} & N_{22} \\
            \vdots                &\vdots               &\vdots            &\vdots      &\vdots \\
            \overline{\kappa}_{ij} & \overline{\tau}_{ij} & \overline{a}_{ij} & \sigma_{ij} & N_{ij}
        \end{pmatrix}
    },
    \onedarray{y}_j = {
        \begin{pmatrix}
            y_{11} \\
            y_{12} \\
            \vdots \\
            y_{21} \\
            y_{22} \\
            \vdots \\
            y_{ij}
        \end{pmatrix}
    }
\end{align}
The \gls{sl} uses the \gls{sil} score based on the median value $ \hat{x}_{{d}_{ij}}$ of a subsequences original feature space per feature $d$.
\begin{align}
    \glsmes{sl} &= f(\widetilde{X}_{\glsmes{sl}}, \onedarray{y}_j) \label{eq:silmedian}
\end{align}
with
\begin{align}
    \widetilde{X}_{\glsmes{sl}} = {
        \begin{pmatrix}
            \hat{x}_{1_{11}} & \hat{x}_{2_{11}}& \hdots & \hat{x}_{d_{11}}& \sigma_{11} & N_{11} \\
            \hat{x}_{1_{12}} & \hat{x}_{2_{12}}& \hdots & \hat{x}_{d_{12}}& \sigma_{12} & N_{12} \\
            \vdots         & & & & \vdots      &\vdots  \\
            \hat{x}_{1_{21}} & \hat{x}_{2_{21}}& \hdots & \hat{x}_{d_{21}}& \sigma_{21} & N_{21} \\
            \hat{x}_{1_{22}} & \hat{x}_{2_{22}}& \hdots & \hat{x}_{d_{22}}& \sigma_{22} & N_{22} \\
            \vdots         & & & & \vdots      &\vdots  \\
            \hat{x}_{1_{ij}} & \hat{x}_{2_{ij}}& \hdots & \hat{x}_{d_{ij}}& \sigma_{ij} & N_{ij}
        \end{pmatrix}
    }
\end{align}
The main idea of this approach is to combine three main parts inside one metric.
First incentive is to reward a \textbf{low standard deviation of the curve parameters} in between a cluster (accomplished by \gls{cc}).
Second, to benchmark the clusters \textbf{spatial separation based on the new feature space} (curve parameters, accomplished by \gls{sp}).
And third, to benchmark the clusters \textbf{spatial separation based on the median of the subsequence in the original feature space} (accomplished by \gls{sl}).
The proposed algorithm for this new metrics computation is described in~\cref{algo:mt3scm}.
\begin{algorithm}[h!] 
    \caption{\gls{mt3scm}} 
    \label{algo:mt3scm} 
    \begin{algorithmic}[1] 
        \Procedure{MT3SCM}{$X,y$}\Comment{Data $X \in \mathbb{R}^{d\times n}$ and labels $\onedarray{y} \in \mathbb{N}^{n}$}
        \State $L \gets empty()$ \Comment{Array initialization for all subsequence median coordinates or \textbf{L}ocation}
        \State $P \gets empty()$ \Comment{Array initialization for all subsequence curve \textbf{P}arameters mean values}
        \State $K \gets$ \Call{GetCurveParametersForAllData}{$X$}
        \State $\onedarray{y}_{unique} \gets$ \Call{FindUniqueClusterIDs}{$\onedarray{y}$}
        \For{$i$ in $\onedarray{y}_{unique}$} 
            \State $X_{i} \gets$ \Call{GetClusterData}{$X,i$}
            \State $\onedarray{s} \gets$ \Call{FindSubsequences}{$y,i$}
            \For{$j$ in $\onedarray{s}$}
                \State $X_{i,j} \gets$ \Call{GetSubsequenceData}{$X_{i},j$}\label{algo:getsubsequencedata}
                \State $L[i, j] \gets$ \Call{GetMedianLocations}{$X_{i,j}$}
                \State $P[i, j] \gets$ \Call{GetCurveParameterValues}{$K,i,j$}
            \EndFor
            \State $cc_i \gets$ \Call{ClusterCurvatureConsistency}{$P$}\label{algo:ClusterCurvatureConsistency} \Comment{Compute the cluster curvature consistency ($cc_i$) with the empirical standard deviation of each curve parameter over time. If the cluster consists only of one time step, set the $cc_i$ to zero.}
            \State $C[i] \gets cc_i$ \Comment{Collect $cc_i$ data for all clusters}
            \EndFor
        \State $\glsmes{wcc} \gets$ \Call{WeightedAverage}{$C, n_{pc}$} \Comment{Compute weighted average curvature consistency (\glsmes{wcc}) from $cc_i$ with number of points per cluster}
        \State $\glsmes{sl} \gets$ \Call{SilhouetteComputation}{$L,\onedarray{y}_{unique}$}\label{algo:mt3scm-silhouette-centers} \Comment{Compute the \gls{sil} coefficient using the center positions of each identified subsequence}
        \State $\glsmes{sp} \gets$ \Call{SilhouetteComputation}{$P,\onedarray{y}_{unique}$}\label{algo:mt3scm-silhouette-features} \Comment{Compute the \gls{sil} coefficient with the curve parameters}
        \State $score \gets (\glsmes{wcc} + \glsmes{sl} + \glsmes{sp}) / 3$
        \State \Return $score$\Comment{The final score}
        \EndProcedure
    \end{algorithmic}
\end{algorithm}

\subsection{Evaluation}
\begin{figure*}[htb]
    \centering
    \includegraphics[width=\textwidth]{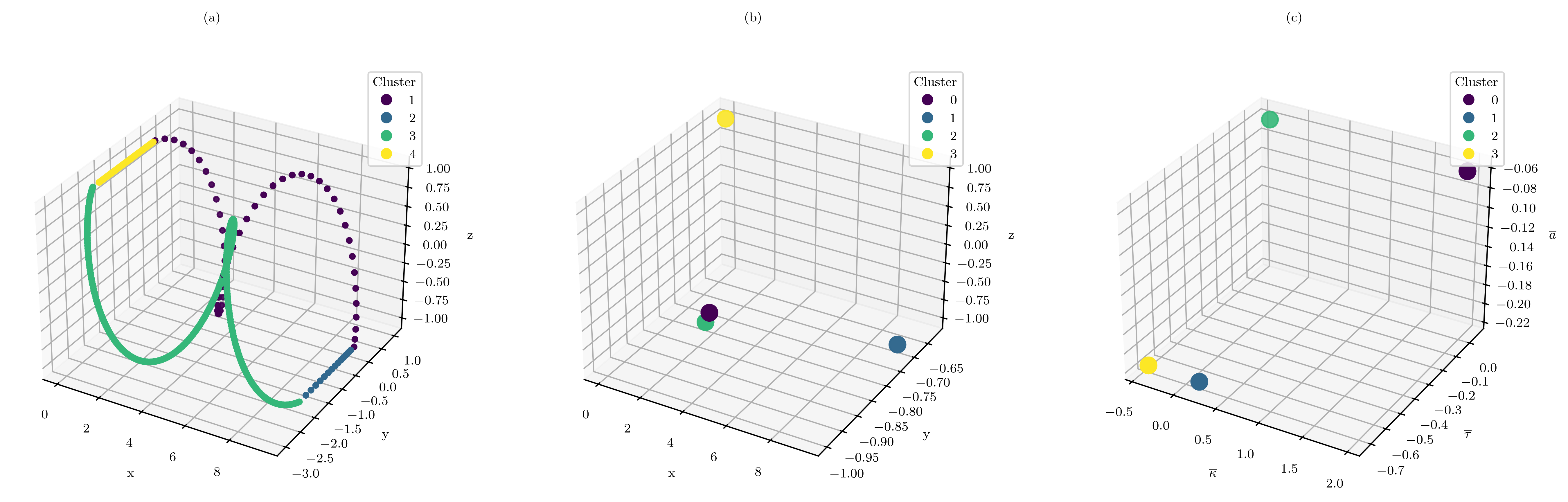}
    \caption{Synthetic dataset with four clusters with a perfect own metric score of $mt3scm=1$ due to each cluster's unique and constant curve parameters. (a) Synthetic dataset with best own result of $mt3scm = 1$. Standard metrics scores computed with original data; \gls{db}: $1.4$, \gls{ch}: $6.9e+02$, \gls{sil}: $0.087$. (b) New feature space from the centers (median value) of each subsequence. Standard metrics scores computed with new feature space; \gls{db}: $6.3e-07$, \gls{ch}: $3.8e+14$, \gls{sil}: $1$. (c) New feature space from the curve parameters extracted from each subsequence. Standard metrics scores computed with new feature space; \gls{db}:~$6.8e-07$, \gls{ch}:~$1.2e+13$, \gls{sil}:~$1$}
    \label{fig:metric-feature-space}
\end{figure*}

\begin{figure}[htb]
    \centering
    \includegraphics[width=.5\textwidth]{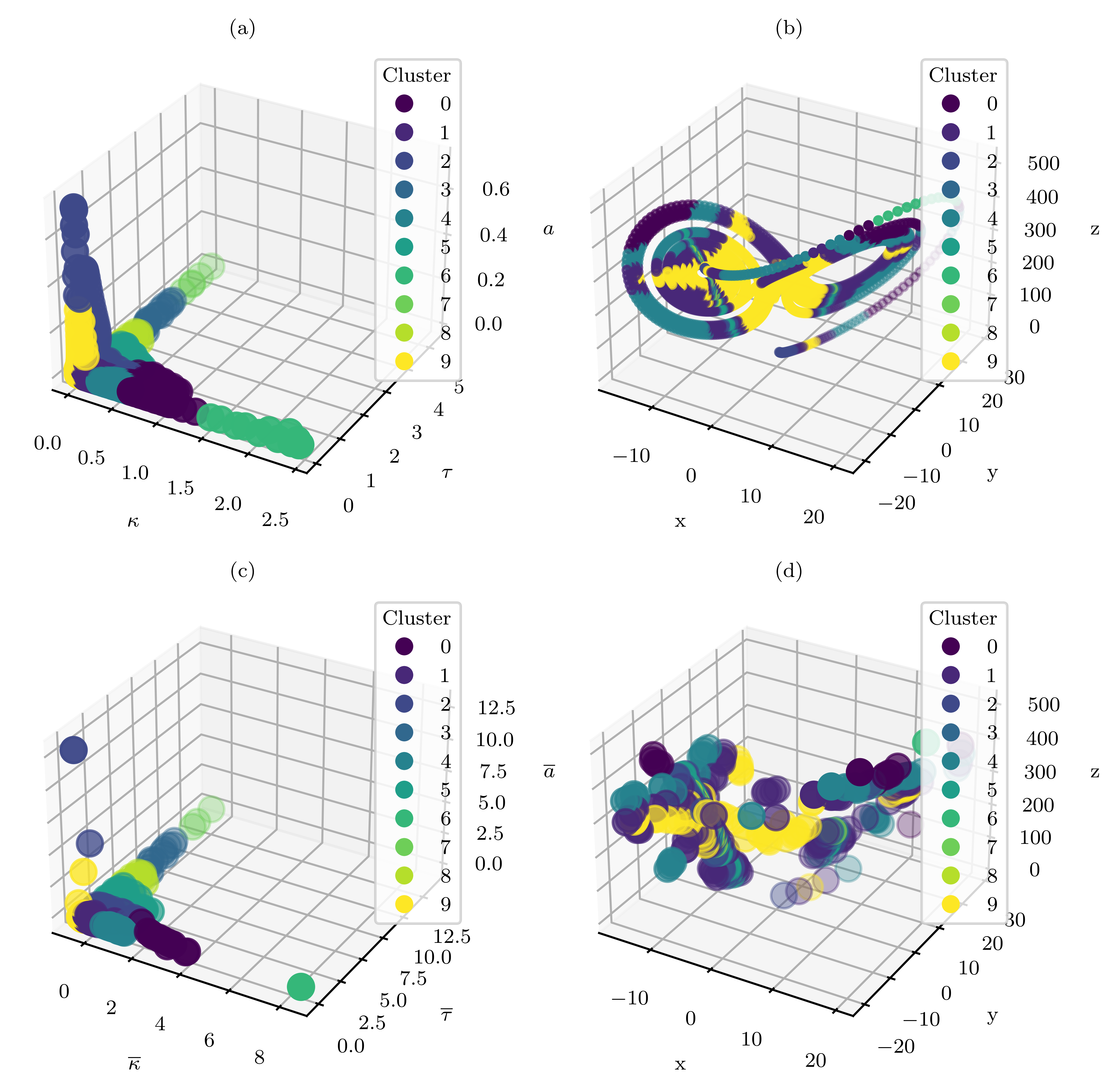}
    \caption{Lorenz-attractor dataset with 10 clusters from agglomerative clustering on the new curve parameters feature space. See \cref{tab:metrics-for-plots} for metric comparison of the following subplots. (a) New curve parameters feature space computed from the Lorenz-attractor dataset with labels from agglomerative clustering (b) Lorenz-attractor dataset with labels from agglomerative clustering on the new curve parameters feature space (c) New feature space from the curve parameters extracted from each subsequence. Standard metrics scores computed with new feature space (d) New feature space from the centers (median value) of each subsequence. Standard metrics scores computed with new feature space}
    \label{fig:metric-lorenz-feature-space}
\end{figure}

For computational tests, we manually created a \myemph{perfect} synthetic dataset with respect to our metric (see \cref{fig:metric-feature-space}).
\Cref{fig:metric-feature-space} (a) shows the original synthetic dataset, where the subsequences in cluster $1$ are a helix along the increasing $x$ axis.
For cluster $2$ the subsequences are a straight movement, with quadratic decreasing distances along the $y$ axis.
Cluster $3$ is representing a helix along the decreasing $x$ axis but with a different resolution than cluster $1$.
Cluster $4$ is, along with cluster $2$, a straight movement with quadratic increasing distances along the $y$ axis.
This cycle is repeated six times.
\Cref{fig:metric-feature-space} (b) shows the new feature space for the \gls{sl} component.
The feature space for the \gls{sp} component is shown in~\cref{fig:metric-feature-space} (c).
Applying the new features per subsequence on the standard metrics, results in best scores for all metrics.
This shows that the new feature space allows a good separation in contrast to the original space, as proven by the metrics scores for \gls{sil}, \gls{ch} and \gls{db} on the original and the two new feature spaces.
To show the benefit of the new feature space, we applied the agglomerative clustering\footnote{\url{https://scikit-learn.org/stable/modules/generated/sklearn.cluster.AgglomerativeClustering.html}} not on the original lorenz-attractor dataset but on the newly computed feature space based on curvature, torsion and acceleration (see \cref{fig:metric-lorenz-feature-space})
The metric values for \cref{fig:metric-lorenz-feature-space} (b) show a high \gls{wcc} and a decent \gls{sp} value for the low number of $10$ clusters specified.

\begin{figure}
    \centering
    \includegraphics[width=.5\textwidth]{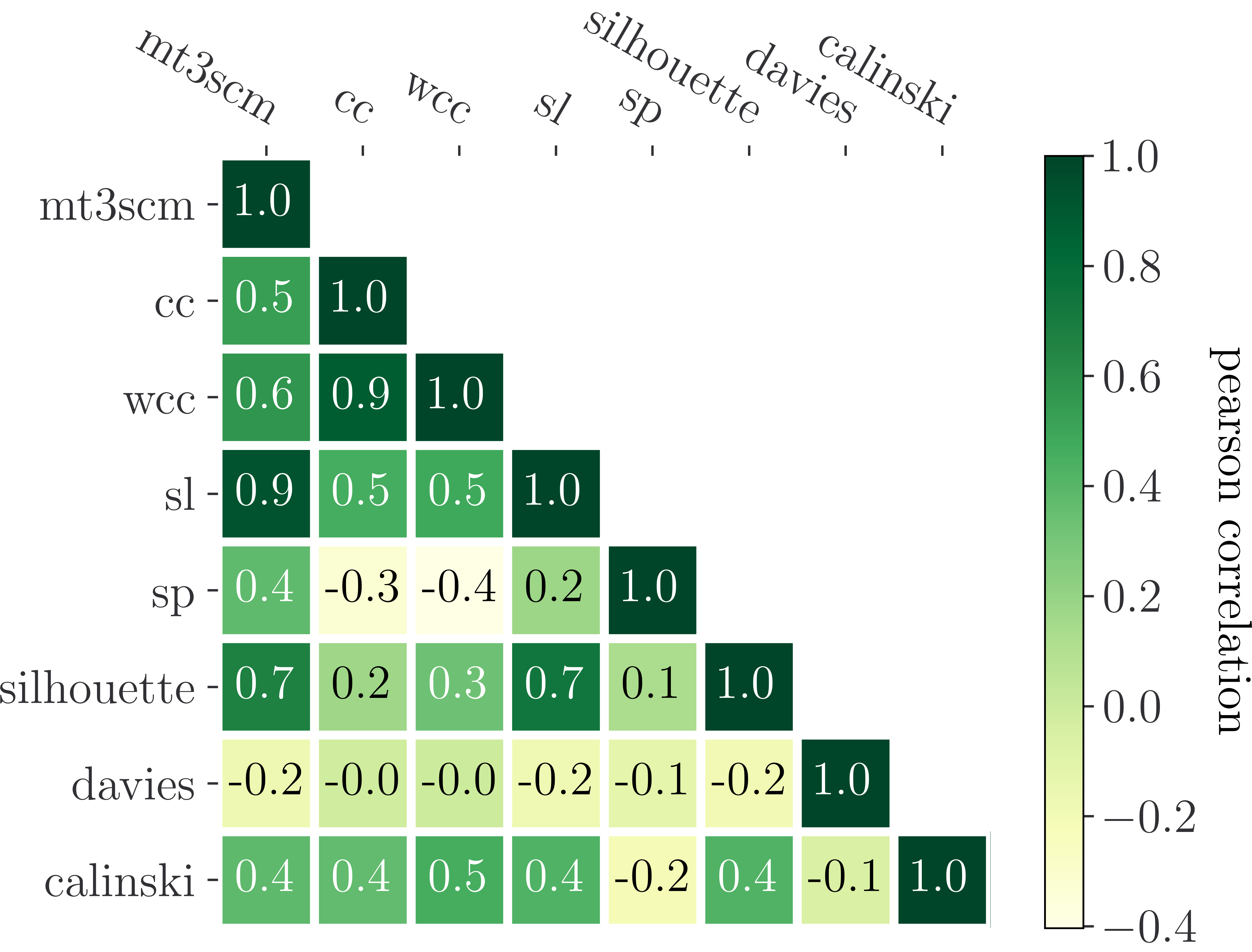}
    \caption{Own metric (\gls{mt3scm}) correlation analysis. Own metric and its four subcomponents (\glsfirst{cc}, \glsfirst{wcc}, \glsfirst{sl}, \glsfirst{sp}) correlation to \gls{ch}, \gls{db} and \gls{sil} score for random, agglomerative and \gls{kmeans} clustering on lorenz and thomas-attractor dataset.}
    \label{fig:metric-correlation}
\end{figure}

\begin{figure}
    \centering
    \includegraphics[width=.5\textwidth]{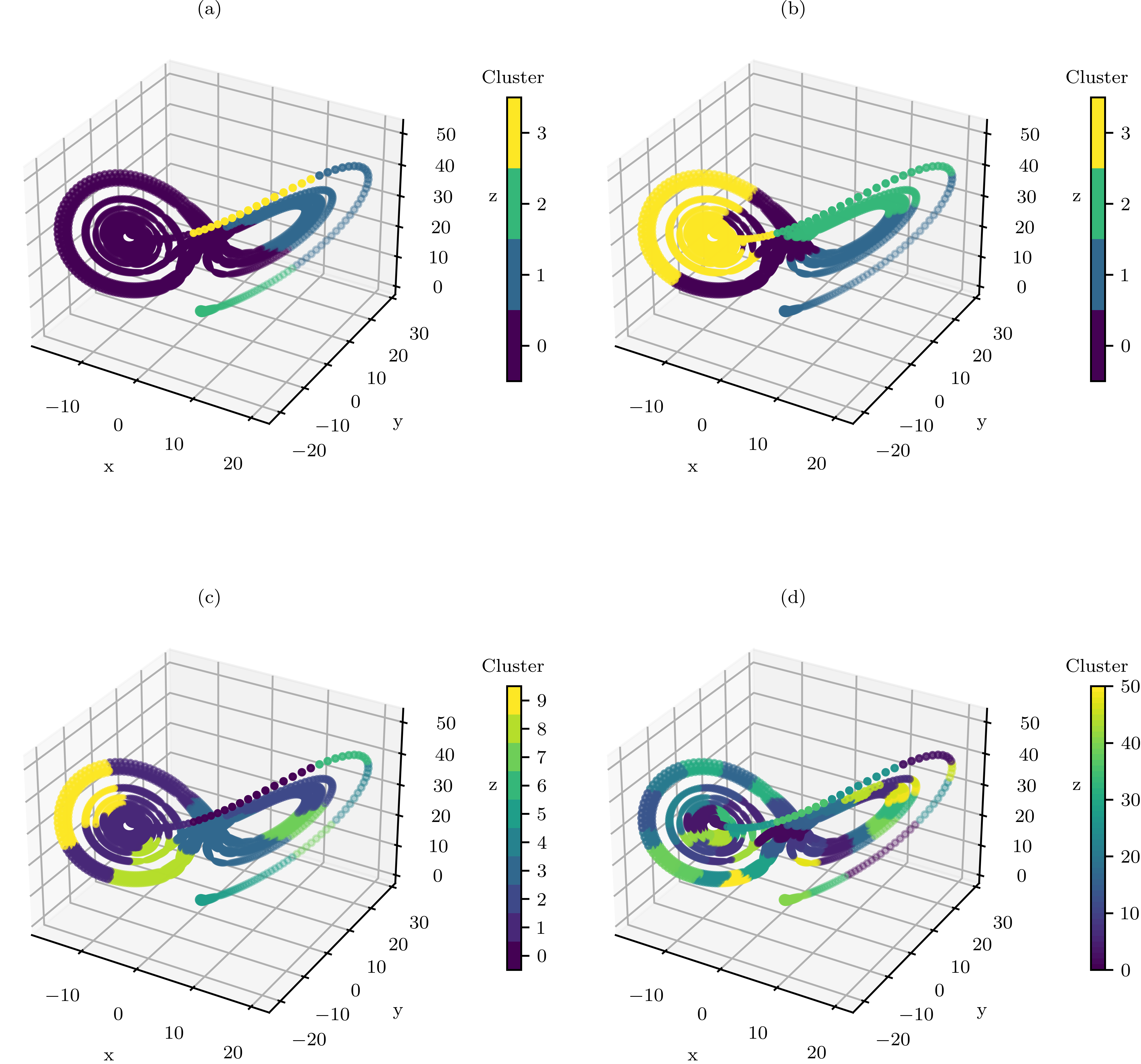}
    \caption{Agglomerative clustering from~\cite{Pedregosa.2011} applied on the lorenz-attractor dataset exemplifies the unique components of our metric compared to the \gls{sil} \gls{ch} and \gls{db} scores. See \cref{tab:metrics-for-plots} for metric comparison of the following subplots. Subfigures (a) (b) and (c) all have a similarly low \gls{mt3scm} score compared to \cref{fig:metric-lorenz-feature-space} (b) but considerably good standard metric scores. Subfigure (d) can achieve a relatively high \gls{mt3scm} score due to the high number of clusters and the resulting good \gls{wcc} and \gls{sl} value which compensates the low \gls{sp} value.}
    \label{fig:metric-agglomerative}
\end{figure}

\begin{figure}
    \centering
    \includegraphics[width=.5\textwidth]{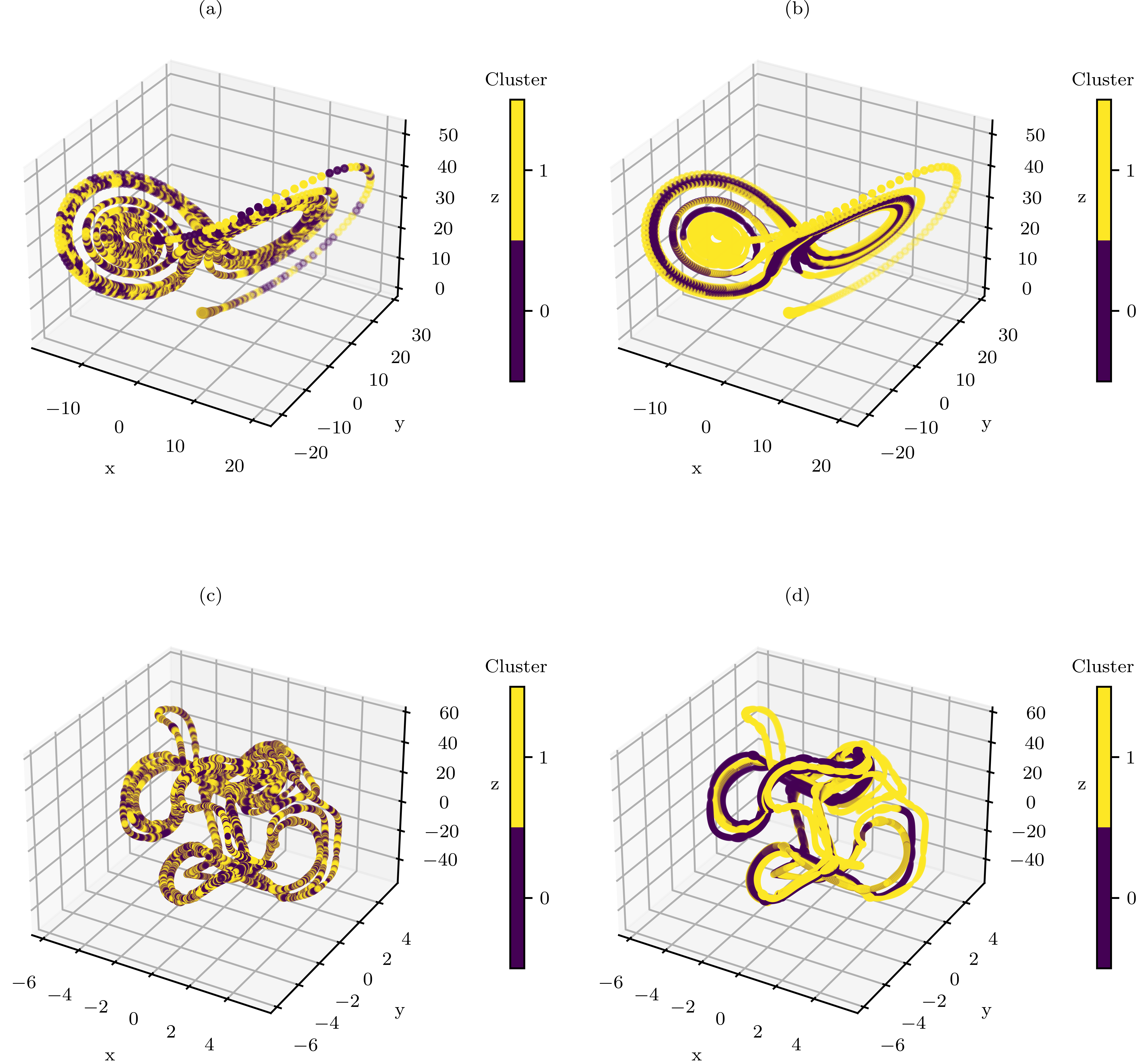}
    \caption{Own metric evaluation using random clusterer on thomas-attractor dataset and lorenz-attractor dataset. See \cref{tab:metrics-for-plots} for metric comparison of the following subplots (a) Own metric and all of its subcomponents are around zero, as desired. \Gls{ch} value is low and \gls{db} is high, which also indicate a \myemph{bad} clustering (b) Longer random subsequences also generate a \gls{mt3scm} result around zero. \Gls{ch} and \gls{db} scores are stronger influenced by the subsequence length (c) Own metric and all of its subcomponents are around zero, as desired. \Gls{ch} value is low and \gls{db} is high, which also indicate a \myemph{bad} clustering (d) As seen for the lorenz-attractor data in~(b), longer subsequences have a high impact on \gls{ch} and \gls{db} scores}
    \label{fig:metric-random}
\end{figure}

To further evaluate our metric, we used the lorenz-attractor and the thomas-attractor dataset (see \cref{tab:datasets}) and applied an agglomerative clustering, a \gls{ts} \gls{kmeans} clustering as well as a random subsequence clustering.
Varying the number of clusters and some algorithm specific parameters.
Afterwards the metrics \gls{ch}, \gls{db} and \gls{sil} scores were computed and compared to our new metric \gls{mt3scm}.
From these results we derived a correlation matrix (see \cref{fig:metric-correlation}).
The \gls{cc} and the \gls{wcc} are clearly related due to their direct combination.
The positive correlation between the internal components to the overall \gls{mt3scm} score is obvious.
We see a clear positive correlation to the \gls{sil} score which is evident due to the internal use of this metric.
Interestingly, the correlation between the \gls{wcc} and the \gls{sp} is negative.
This is due to the types of datasets and algorithms we used.
Because with higher number of clusters we theoretically expect a better \gls{cc} because of the lower standard deviation by chance.
On the other hand, the more clusters exist, the more likely a similar curve parameter between the clusters exists and therefore creates a new feature space with overlapping clusters, which results in a low \gls{sp} score.
This can be retraced within the subfigures of \cref{fig:metric-agglomerative}.
The low correlation between the \gls{ch} and the \gls{db} scores supports our point that the available clustering metrics are not well suited to be used for \gls{ts} clustering evaluation measures.
\begin{table*}[htbp]
    \small
    \centering
    \caption{Metric values for \cref{fig:metric-agglomerative,fig:metric-random,fig:metric-lorenz-feature-space}}
    \label{tab:metrics-for-plots}
    \begin{tabularx}{\textwidth}{XXXXXXXXX}
    \toprule
    Figure & \gls{db} & \gls{ch} & \gls{sil} & \gls{mt3scm} & $cc$ & $cc_w$ & \gls{sl} & \gls{sp}\\
    \midrule
    \cref{fig:metric-lorenz-feature-space}~(a) & 0.64 & 4.5e+03 & 0.51 & 0.1 & 0.51 & 0.29 &  0.045 & -0.026\\
    \cref{fig:metric-lorenz-feature-space}~(b) & 13 & 86 & -0.32 & 0.23 & 0.65 & 0.77 & -0.19 & 0.11\\
    \cref{fig:metric-lorenz-feature-space}~(c) & 0.71 & 2.3e+02 & 0.43 & 0.13 & 0.4 & 0.4 & 0 & 0\\
    \cref{fig:metric-lorenz-feature-space}~(d) &  63 & 9.1 & -0.3 & 0.18 & 0.36 & 0.54 & 0 & 0\\
    \midrule
    \cref{fig:metric-agglomerative}~(a) & 0.74  & 6.6e+02 & 0.34 & 0.071 & 0.24 & 0.22 & 0.02 & -0.029\\
    \cref{fig:metric-agglomerative}~(b) & 1.1 & 1.9e+03 & 0.32 & 0.027 & 0.25 & 0.23 & -0.013 & -0.13 \\
    \cref{fig:metric-agglomerative}~(c) & 0.85 & 1.1e+03 & 0.29 & 0.034 & 0.54 & 0.42 & -0.042 & -0.28 \\
    \cref{fig:metric-agglomerative}~(d) &  0.78 & 2.4e+03 & 0.33 & 0.26 & 0.78 & 0.73 & 0.37 & -0.33\\
    \midrule
    \cref{fig:metric-random}~(a) & 70 & 0.49 & 3.4E-05 & -0.00092 & -0.00096 & -0.00089 & -0.0012 & -0.00068 \\
    \cref{fig:metric-random}~(b) & 7 & 48 & 0.029 & -0.053 & 0.092 & 0.078 & -0.12 & -0.12 \\
    \cref{fig:metric-random}~(c) & 65 & 1.6 & 5.2E-05 & -0.00047 & -0.00084 & -0.00083 & -0.00037 & -0.00022 \\
    \cref{fig:metric-random}~(d) & 3.5 & 5.4E+02 & 0.039 & -0.0035 & 0.00041 & 0.0011 & -0.0025 & -0.0091 \\
     \bottomrule
    \end{tabularx}
\end{table*}
\Cref{fig:metric-agglomerative} shows examples where the agglomerative clustering was applied on the lorenz-attractor dataset (part of the data used for the correlation matrix \cref{fig:metric-correlation}).
It can be seen that the agglomerative clustering on the original dataset is not an optimal cluster algorithm, when comparing the metrics to \cref{fig:metric-lorenz-feature-space} (b).
Comparing \cref{fig:metric-lorenz-feature-space} (b) and \cref{fig:metric-agglomerative} (d) we can see a similar \gls{mt3scm} score but very different standard metrics scores. The similar \gls{mt3scm} score is based on the much higher number of clusters and equally distributed subsequence length in \cref{fig:metric-agglomerative} (d), which results in a high \gls{wcc} value as well as a good spatial separation (\gls{sl}), which is compensating the low \gls{sp} value due to the similar curve parameters of the clusters. \cref{fig:metric-lorenz-feature-space} (b) however, also has a very high \gls{wcc} value with a good \gls{sp} value reaching a similar \gls{mt3scm} score but with a fifth of the number of clusters.
How our metric handles random clustering with critical scenarios, is shown in \cref{fig:metric-random}.
The Python code and a more detailed evaluation are publicly available at~\cite{Kohne.2022}

\subsection{Conclusion}
We have described a more suitable similarity measure for dependent \gls{tsd}.
After showing how to compute our metric and evaluated on different datasets its use case and effectiveness.
Further we will use this metric in addition to the standard metrics to evaluate our proposed online \gls{ts} clustering algorithm which is described in \vref{sec:Aglorithm}

\section{Clustering Algorithm (ABIMCA)}\label{sec:Aglorithm}
In this section we describe the concept of our \gls{ts} clustering approach in detail. Afterwards, we apply our algorithm onto the datasets described in \vref{sec:Datasets} and present the results.

\subsection{Method}
As described in \cite{Liao.2005} a key component in a \gls{ts} clustering algorithm is the similarity function to quantify the clustering criteria.
Common similarity functions used are distance measures like euclidean distance or some kind of correlation coefficients like Pearson's correlation coefficient.
Those are also used for static data clustering algorithms.
More suitable for \gls{ts} clustering are similarity functions like \gls{dtw} distance, \gls{sts} distance~\cite{MollerLevet.2003} or considering space curves like we introduced in~\cref{sec:Metric}.
\begin{figure}[!h]
    \centering
    \includegraphics{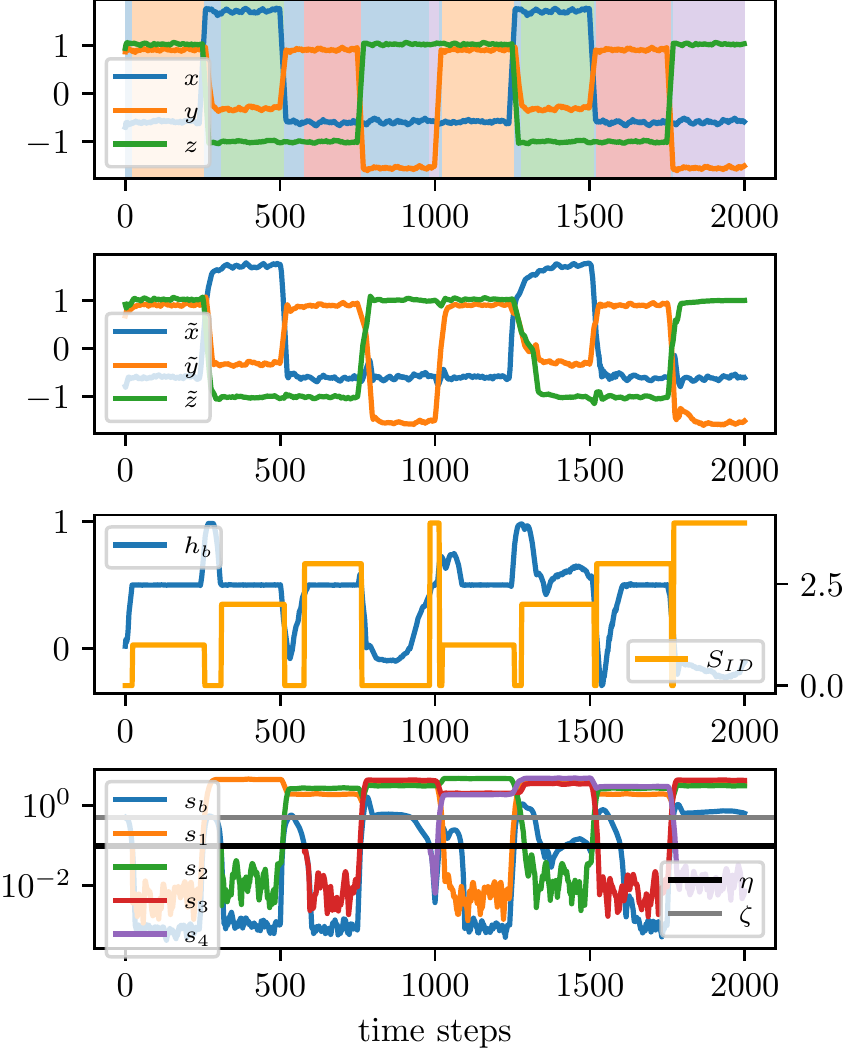}
    \caption{Example of online clustering with a simple three-dimensional synthetic dataset. First row shows the original input data $X$ (or the last values of each sliding window $W_t$) with the online cluster IDs as the background color (blue is unknown or $S_{ID} = 0)$. Second row shows the output of the \gls{ae} or the reconstruction $\widetilde{W}$. In the third row the blue line represents the value of the latent space $h_b$ (left axis) and the identified subsequence ID $S_{ID}$ (right axis). The last row indicates the \gls{bae}['s] ($s_b$) as well as the \glsplural{sae}['] ($s_1$ - $s_4$) score values. The black horizontal line is the subsequence detection score threshold ($\eta$) and the gray line is the subsequence recognition score threshold ($\rho$).}
    \label{fig:training-score-example}
\end{figure}
\begin{figure}[h]
    \centering
    \includegraphics{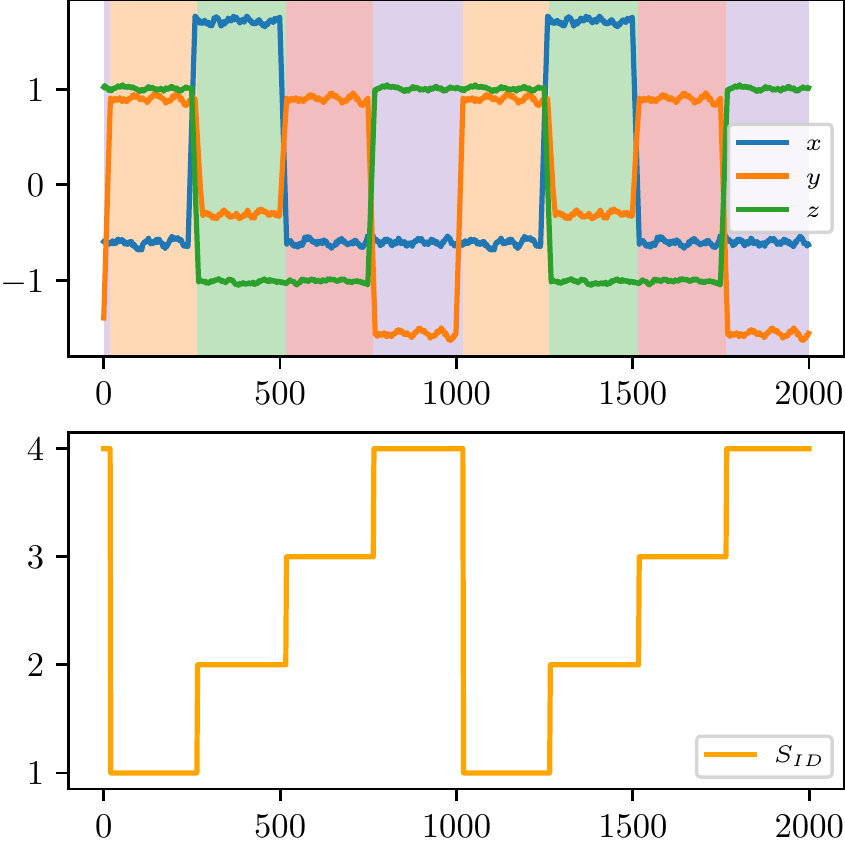}
    \caption{Example of batch-wise offline clustering with a simple three-dimensional synthetic dataset.}
    \label{fig:prediction-score-example}
\end{figure}

\begin{figure*}[h!]
    \centering
    \def\svgwidth{\columnwidth}
\begingroup%
  \makeatletter%
  \providecommand\color[2][]{%
    \errmessage{(Inkscape) Color is used for the text in Inkscape, but the package 'color.sty' is not loaded}%
    \renewcommand\color[2][]{}%
  }%
  \providecommand\transparent[1]{%
    \errmessage{(Inkscape) Transparency is used (non-zero) for the text in Inkscape, but the package 'transparent.sty' is not loaded}%
    \renewcommand\transparent[1]{}%
  }%
  \providecommand\rotatebox[2]{#2}%
  \newcommand*\fsize{\dimexpr\f@size pt\relax}%
  \newcommand*\lineheight[1]{\fontsize{\fsize}{#1\fsize}\selectfont}%
  \ifx\svgwidth\undefined%
    \setlength{\unitlength}{282.29513171bp}%
    \ifx\svgscale\undefined%
      \relax%
    \else%
      \setlength{\unitlength}{\unitlength * \real{\svgscale}}%
    \fi%
  \else%
    \setlength{\unitlength}{\svgwidth}%
  \fi%
  \global\let\svgwidth\undefined%
  \global\let\svgscale\undefined%
  \makeatother%
  \begin{picture}(1,0.48386136)%
    \lineheight{1}%
    \setlength\tabcolsep{0pt}%
    \put(0.47386007,0.00184474){\makebox(0,0)[t]{\lineheight{1.25}\smash{\begin{tabular}[t]{c}$\mathcal{L}(x, h, r)$\end{tabular}}}}%
    \put(0.38435354,0.06881444){\makebox(0,0)[t]{\lineheight{1.25}\smash{\begin{tabular}[t]{c}Base AE\end{tabular}}}}%
    \put(0.61746658,0.09598418){\makebox(0,0)[t]{\lineheight{1.25}\smash{\begin{tabular}[t]{c}$\mathcal{L}_{C_i}$\end{tabular}}}}%
    \put(0.16132175,0.34615475){\makebox(0,0)[t]{\lineheight{1.25}\smash{\begin{tabular}[t]{c}$W_t$\end{tabular}}}}%
    \put(0.68873886,0.06215121){\makebox(0,0)[t]{\lineheight{1.25}\smash{\begin{tabular}[t]{c}$\mathcal{L}$\end{tabular}}}}%
    \put(0,0){\includegraphics[width=\unitlength,page=1]{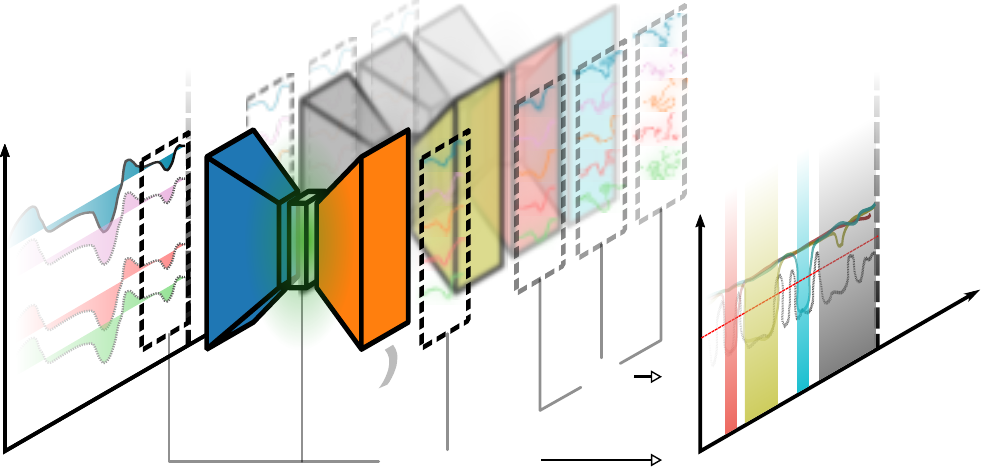}}%
    \put(0.19111883,0.10926544){\makebox(0,0)[t]{\lineheight{1.25}\smash{\begin{tabular}[t]{c}$t$\end{tabular}}}}%
    \put(0.89694967,0.10926544){\makebox(0,0)[t]{\lineheight{1.25}\smash{\begin{tabular}[t]{c}$t$\end{tabular}}}}%
  \end{picture}%
\endgroup%

    \caption{Concept of the ABIMCA approach. Sliding window of the \gls{mts} $W_t$ is iteratively trained in the base \gls{ae}. If score of base \gls{ae} (gray dotted line) is below threshold (dashed red line), a new subsequence \gls{ae} is created from the base \gls{ae}. Incoming data is also compared to existing subsequence \glsplural{ae} if subsequence can be recognized.}
    \label{fig:abimca-concept}
\end{figure*}

In this work we analyzed an approach which is data driven, based on unsupervised \gls{ml} algorithms and has online capabilities (see~\cref{fig:abimca-concept}).
Our approach uses a \gls{RNN} based \gls{ae} to generate scores which are used as similarity measures.
Specifically, the experiments in this work were performed using a pytorch~\cite{Paszke.2019} implementation of a bidirectional one-layer \gls{GRU} \gls{RNN} with a hidden size of the input dimensions minus one $h=d-1$.
Other prerequisites regarding the dataset and preprocessing are described in \vref{sec:Datasets} and \vref{sec:Definitions-and-Restrictions}.

The main procedure of the approach is as follows:
The incoming data is taken as a sliding window $W_t$ at the current time $t$ with length $\zeta$ of past time steps and number of features $d$.
This matrix $W_t \in \mathbb{R}^{d\times\zeta}$ is used for the input of, what we call, the \glsfirst{bae}. The key element of our algorithm is, that this \gls{bae}['s] parameters are not constant but being adapted iteratively with a \gls{sgd} optimization method for each new incoming sliding window.
For this training of the \gls{bae}, we use a slight adaption of the sparse \gls{ae} loss function~$\mathcal{L}$ from \cite{Ranzato.2006} with a basic regularization term or sparsity penalty~$\Omega$
\begin{align}
    loss = l = \lossfun &= MSE(W_t, \widetilde{W}_t) + \Omega(\latspace)\label{eq:lossfunction1}
\end{align}
where $\latspace = f(\mathbf{x})$ is the encoders output or latent space. The sparsity penalty we denote as:
\begin{align}
    \Omega(\latspace) &= \lambda \cdot \sum_{i}^{d-1} |h_i - c_{lc}|\label{eq:sparsitypenalty}
\end{align}
with the penalty factor $\lambda = 1\mathrm{e}{-10}$ and the latent center constant $c_{lc} = 0.5$.
The \gls{mse} is
\begin{align}
    MSE(W_t, \widetilde{W}_t) &= \frac {1}{d\cdot \zeta}\sum_{i=1}^d \sum_{j=1}^\zeta (w_{ij} - \widetilde{w}_{ij})^2
\end{align}
which results in the final loss computation
\begin{align}
    loss = l &= \frac {1}{d\cdot \zeta}\sum_{i=1}^d \sum_{j=1}^\zeta (w_{ij} - \widetilde{w}_{ij})^2 + \lambda \cdot \sum_{i}^{d-1} |h_i - c_{lc}|\label{eq:lossfunction2}
\end{align}
where the first part is the \gls{mse} between the input matrix $W_t$ and the reconstruction $\widetilde{W}_t$ and the second part is the penalty of the latent space deviation.

To determine if a subsequence is recognized at the current time step, we denote the scoring function $SF$ as follows
\begin{align}
    s = score = SF(l, \latspace) &=  c_{fw}\cdot(|c_{lc} - \latspace|) + \frac{l}{c_{fw}}\label{eq:scoringfunction}
\end{align}
whereas the weighting factor in current implementation is $c_{fw} = 1$ and latent center constant is $c_{lc} = 0.5$.
It utilizes the reconstruction error as well as the deviation of the latent space with a doubled emphasis on the latent space deviation due to its dependence in the loss function as well as the scoring function which includes the loss again (see \cref{eq:scoringfunction}).
In combination with a threshold, the score is used to determine when a recognizable subsequence is present.

The bottom row of \cref{fig:training-score-example} shows, that a subsequence is present, when the \gls{bae}['s] score (blue line) is below the horizontal black line ($s_b <= \eta$).
If a subsequence is present, a copy of the \gls{bae} is made and its parameters are frozen and associated with this specific pattern of a subsequence.
These copies of the \gls{bae}, which we call \glsfirst{sae}, are used to recognize previously seen subsequences using the same scoring function.
A concept drawing of the approach is shown in~\cref{fig:abimca-concept}.
The algorithm is described in pseudocode in \cref{algo:abimca}.

The functionality can be retraced considering \cref{fig:training-score-example,fig:prediction-score-example}.
This example shows the algorithm applied on a three-dimensional synthetic data set.
The input data consists of four different operation points with small white noise.
The sequence of the four subsequences is repeated once.
The other rows are described in the caption of \cref{fig:training-score-example}.
It can be seen that the algorithm needs a few time steps to adapt to the current subsequence until it is recognized as such.
Recognizing a previously identified subsequence, however, is almost instantaneous.
It is apparent, how the calibration of the thresholds $\eta$ (horizontal black line) and $\zeta$ (horizontal gray line) are crucial.
The necessary time steps to adapt to a current subsequence can be altered by the calibration of the learning rate $\alpha$ and the number of \gls{bae}['s] training cycles per time step $\omega$.
A faster recognition of a subsequence has the drawback of the algorithm being very sensitive and therefore identifying even small changes of the input as a new subsequence.
A strategy could be to calibrate the algorithm first to be rather insensitive and cluster the \gls{ts} in major subsequences.
These can then be further clustered with a more sensitive calibration.
This procedure can be repeated until the required degree of granularity is achieved.
For a streaming application multiple runs of this procedure could also be applied in parallel and combined into a cluster tree.

\begin{algorithm}[h!] 
    \caption{\gls{abimca}. Using the following parameters: $\alpha$: learning rate, $\omega$: number of base model training cycle per step, $\eta$: subsequence detection score threshold, $\rho$: subsequence recognition score threshold, window length: $\zeta$. Also, we denote $\theta_{S}$ as a list of subsequence model parameters, $\onedarray{s}$ as an array of scores for all existing subsequence models, $\widetilde{W}_t$ the reconstruction of the sliding window input, } 
    \label{algo:abimca} 
    \begin{algorithmic}[1] 
        \Procedure{Abimca}{$W_t$}\Comment{Sliding window data $W_t \in \mathbb{R}^{d\times\zeta}$}
        \State Verify calibration $t \geq \zeta \vee \alpha > 0 \vee \omega \ge 1 \vee \eta > 0 \vee \rho > \eta$
        \State $c_S \gets 0$ \Comment{Initialize subsequence counter}
        \State $\theta_{b} \gets$ \Call{Sparse}{$sparsity=0.1$}\Comment{Initialize base model parameters}
        \While{$W_t$}\Comment{New input available}
            \For{ $j\gets 1,\omega$} \Comment{Iterative base model train loop}
                \State $\widetilde{W}_t, \latspace_b \gets $\Call{Predict}{$W_t, \theta_b$} 
                \State $l_b \gets \lossfun$ 
                \State $\Delta\theta_b \gets $\Call{Backpropagate}{$l_b$}
                \State $\theta_{b} \gets \theta_b + \Delta\theta_b$ \Comment{Update base model parameters}
            \EndFor
            \State $\widetilde{W}_t, \latspace_b \gets $\Call{Predict}{$W_t, \theta_b$}
            \State $s_b \gets $\Call{SF}{$l_b, \latspace_b$} 
            \State $\onedarray{s} \gets $\Call{GetSubsequenceScores}{$W_t, \mathbf{\theta_{S}}$} 
            \If{$\min (\onedarray{s}) < \rho$} \Comment{Subsequence recognized}
                \State{$S_{ID} \gets \argmin (\onedarray{s})$} \Comment{Set ID to recognized subsequence index}
            \Else{}  \Comment{No subsequence recognized}
                \If{$s_b <= \eta$} \Comment{Score below new subsequence threshold}
                    \State $\mathbf{\theta_{S}}.append(\theta_{b})$  \Comment{Append current base model parameters to list of subsequence models parameters}
                    \State $c_s \gets c_s + 1$ \Comment{increase subsequence counter}
                    \State{$S_{ID} \gets c_s$} \Comment{Set ID to new subsequence index}
                \Else \Comment{In transition}
                    \State{$S_{ID} \gets 0$} \Comment{Set ID to unknown}
                \EndIf
            \EndIf
        \EndWhile\label{algo:abimcaendmainwhile}
        \EndProcedure
    \end{algorithmic}
\end{algorithm}

\subsection{Evaluation}
For the evaluation study of our algorithm, we chose eight different \gls{mts} datasets (see \cref{tab:datasets}) from which six are publicly available and two are provided with our codebase \cite{Kohne.2022b}, seven other state-of-the-art algorithms (see \cref{tab:algorithms}) and three widely used unsupervised clustering metrics (see \cref{tab:metrics}).
Each algorithm has been applied to each dataset with default parameters.
Additionally, we performed a hyperparameter search for each algorithm based on a random grid search of 300 samples.
The parameter boundaries for this hyperparameter search are listed in~\cref{tab:hp_search_params}.
Overall, $19\,264$ experiments were run.

\begin{table}[h]
    \small
    \centering
    \caption{Summation of the number of outperformances of each algorithm for all datasets and all metrics compared to the mini-batch-kmeans with parameters from hyperparameter search}
    \label{tab:metrics_outperformance_mini-batch-kmeans}
    \begin{tabular}{lp{.9cm}p{.9cm}p{.9cm}p{.9cm}r}
    \toprule
    Outperforms & \multicolumn{4}{c}{mini-batch-kmeans} & Total \\
    \midrule
    Metric & silhouette & calinski-harabasz & davies-bouldin & mt3scm &  \\
    Algorithm &  &  &  &  &  \\
    \midrule
    CluStream & 0 & 0 & 0 & 3 & 3 \\
    BOCPDetector & 0 & 4 & 2 & 1 & 7 \\
    BIRCH & 2 & 3 & 1 & 2 & 8 \\
    STREAMKMeans & 1 & 2 & 3 & 5 & 11 \\
    DBSTREAM & 2 & \bfseries 6 & 6 & 3 & 17 \\
    DenStream & \bfseries 5 & 4 & \bfseries 8 & 5 & 22 \\
    ABIMCA & \bfseries 5 & 4 & 7 & \bfseries 7 & \bfseries 23 \\
    \bottomrule
    \end{tabular}
\end{table}
\begin{table}[h]
    \small
    \centering
    \caption{Summation of the number of outperformances of each algorithm for all datasets and all metrics compared to the mini-batch-kmeans with default parameters}
    \label{tab:metrics_outperformance_mini-batch-kmeans-default}
    \begin{tabular}{lp{.9cm}p{.9cm}p{.9cm}p{.9cm}r}
    \toprule
    Outperforms & \multicolumn{4}{c}{mini-batch-kmeans} & Total \\
    \midrule
    Metric & silhouette & calinski-harabasz & davies-bouldin & mt3scm &  \\
    Algorithm &  &  &  &  &  \\
    \midrule
    CluStream & 1 & 0 & 0 & 5 & 6 \\
    BIRCH & 3 & 0 & 0 & 5 & 8 \\
    DenStream & 1 & 1 & 3 & \bfseries 6 & 11 \\
    ABIMCA & 3 & 1 & 4 & 4 & 12 \\
    DBSTREAM & 2 & 4 & 3 & 3 & 12 \\
    STREAMKMeans & \bfseries 5 & 1 & 3 & 5 & 14 \\
    BOCPDetector & 1 & \bfseries 6 & \bfseries 6 & 5 & \bfseries 18 \\
    \bottomrule
    \end{tabular}
\end{table}
For a better overview of the results, we chose to compare every algorithm to the \myemph{MiniBatchKMeans} algorithm and counted the number of times they performed better.
\Cref{tab:metrics_outperformance_mini-batch-kmeans} shows the results for the hyperparameter search and the number of outperformances of each algorithm compared to the \myemph{MiniBatchKMeans} algorithm.  \Cref{tab:metrics_outperformance_mini-batch-kmeans-default} shows the same results with default parameter settings for each algorithm.
We can see, that in sum and in two of the metrics our algorithm beats state-of-the-art algorithms.
The full list of results is attached in \vref{tab:default_metrics_metrics.mt3scm,tab:default_metrics_metrics.calinski-harabasz,tab:default_metrics_metrics.davies-bouldin,tab:default_metrics_metrics.silhouette,tab:hp_metrics_metrics.mt3scm,tab:hp_metrics_metrics.calinski-harabasz,tab:hp_metrics_metrics.davies-bouldin,tab:hp_metrics_metrics.silhouette}

As cited before, every algorithm performs differently on the specific distribution of patterns and the hyperparameter search was a simple random grid search of \myemph{only} 300 samples, so these results unlikely represent the optimal solution for each algorithm on each dataset.
Nevertheless, we demonstrate, that the algorithm we present in this work, is highly effective of detecting subsequences online in a \gls{mts}.

\section{Limitations and Discussion}\label{sec:Discussion}
For the evaluation of the segmentation of \gls{mtsd}, we introduced a new metric which is based on space-curve parameters in the feature space.
Due to the wide variety of fields, use cases and applications, this falls into place for some applications and uses cases but not for all.
The calculation of these space-curve parameters is sensitive to outliers and smoothness and questionable for steady-state conditions or \emph{non-moving} point clouds in the feature space.
We have implemented specific numerical boundary limits for computing the derivatives of the data in these states, but it needs to be considered and evaluated if this is compatible with the application.
Because of the outlier sensitivity we use the mean value of these parameters as well as their standard deviation.
A low-pass filter for very noisy data should be considered before applying it to the metric.
Attention is called for, when the data is scaled or standardized.
This effects the actual space curve parameters, since a constant curvature is likely not constant anymore after scaling.
The metric also tends to reward short subsequences who only occur once.
Due to the mean value of the curve parameters the subsequence separation appears to be good, but the variance of one large subsequence is high. This needs to be compensated or prevented more and will be part of future analysis and improvement of the metric.
It might also be reasonable to introduce weighting factors for the three parts of our metric in \cref{eq:mt3scm} to consider domain specific emphasis on a rather spatial or curve parameter separation requirement.

Regarding our clustering method, calibration of the main thresholds ($\eta$, $\rho$) needs special attention.
In combination with the learning rate, they mainly influence the \myemph{sensitivity} of the segmentation process.
Using only the reconstruction error or only the representation deviation can be beneficial for different use cases and complexities (e.g., changing the weighting factor in~\cref{eq:scoringfunction}).
Advantages are that no additional information about the data for offline clustering needs to be saved.
All necessary information for clustering data after training is the parameters of the subsequence specific \glsplural{ae}.
It is a completely unsupervised method which can cluster online data.
In the context of \gls{cbm} the once identified subsequence \glsplural{ae} can be used for deviation quantification of the underlying system.
This can be used for deterioration analysis and maintenance strategies.
Further investigations for improving the \gls{abimca} method would be to explore different kind of \glsplural{ae} like \gls{FNN}, \gls{CNN} or a combination of such.
Also, a \gls{VAE} could be reasonable depending on the underlying process.
Future work should analyze the effect of reducing the latent space dimension by multiple factors of the input dimension (when input dimension is very high).
This could reduce computation costs and improve representation learning without performance loss.
A detailed analysis of the optimal default parameters or a generic automatic calibration depending on some statistics of the expected input could increase performance and decrease calibration efforts.

\section{Conclusion}
\glsresetall
In this paper we have introduced the \gls{abimca} which is a deep learning method to separate \gls{mtsd} into subsequences.
It is beneficial in a variety of fields, to cluster \gls{mtsd} into smaller segments or subsequences in an unsupervised manner.
The ability to filter measurement data based on specific subsequences can improve downstream development products such as anomaly detection or machine diagnosis in \gls{cbm} strategies.
Our algorithm is specifically useful for \gls{mtsd} generated by a \gls{ms} in a transient environment.
It can be used offline as well as online for streaming data.
It utilizes \gls{RNN} based \gls{ae}[s] by iteratively training a \gls{bae}, generating a segmentation score and saving the intermediate parameters of the \gls{bae} to recognize previously identified subsequences.
By comparing our algorithm with seven other algorithms on eight different publicly available datasets using four different unsupervised metrics (from which we introduced one ourselves), we have shown that our algorithm outperforms state-of-the-art algorithms.
Our unsupervised metric introduced (\gls{mt3scm}), is an attempt to use a more intuitive similarity measure based on the curvature and other space-curve parameters of the spanned feature space.
Additionally, all our code is open source and publicly available for benchmarking.
\cleardoublepage

\begin{table*}
    \footnotesize
    \centering
    \caption{Best metric 'metrics.mt3scm' value for each dataset and algorithm from hyperparameter search results}
    \label{tab:hp_metrics_metrics.mt3scm}
        \begin{tabular}{lrrrrrrrr}
        \toprule
        algorithm & birch & bocpdetector & clustream & dbstream & denstream & mini-batch-kmeans & streamkmeans & abimca \\
        dataset &  &  &  &  &  &  &  &  \\
        \midrule
        bee-waggle & 0.139 & 0.143 & 0.125 & 0.051 & 0.183 & 0.075 & 0.099 & \bfseries 0.370 \\
        cmapss & 0.129 & 0.120 & 0.213 & 0.108 & 0.226 & 0.141 & \bfseries 0.378 & 0.179 \\
        eigen-worms & 0.018 & 0.059 & 0.099 & -0.026 & \bfseries 0.389 & 0.064 & 0.099 & 0.150 \\
        hydraulic & 0.364 & 0.104 & 0.055 & 0.006 & 0.590 & 0.655 & \bfseries 0.667 & 0.598 \\
        lorenz-attractor & 0.242 & 0.110 & 0.019 & 0.272 & 0.274 & 0.279 & 0.272 & \bfseries 0.368 \\
        mocap & 0.102 & 0.047 & 0.256 & \bfseries 0.277 & 0.273 & 0.257 & 0.067 & 0.258 \\
        occupancy & 0.430 & 0.214 & 0.084 & \bfseries 0.697 & 0.450 & 0.267 & 0.458 & 0.235 \\
        own-synth & 0.355 & 0.159 & 0.096 & 0.366 & 0.279 & 0.366 & 0.297 & \bfseries 0.622 \\
        thomas-attractor & 0.115 & nan & 0.013 & nan & 0.151 & 0.153 & 0.079 & \bfseries 0.474 \\
        \bottomrule
        \end{tabular}
\end{table*}

\begin{table*}
    \footnotesize
    \centering
    \caption{Best metric 'metrics.calinski-harabasz' value for each dataset and algorithm from hyperparameter search results}
    \label{tab:hp_metrics_metrics.calinski-harabasz}
        \begin{tabular}{lrrrrrrrr}
        \toprule
        algorithm & birch & bocpdetector & clustream & dbstream & denstream & mini-batch-kmeans & streamkmeans & abimca \\
        dataset &  &  &  &  &  &  &  &  \\
        \midrule
        bee-waggle & 225.040 & 169.623 & 6.085 & \bfseries 1193.353 & 357.833 & 210.572 & 297.385 & 324.477 \\
        cmapss & 2.56e+04 & \bfseries 3.84e+04 & 4182.524 & 2731.331 & 9154.261 & 4930.092 & 1244.679 & 7847.877 \\
        eigen-worms & 859.800 & 2498.356 & 94.549 & \bfseries 2945.532 & 1880.220 & 1746.451 & 1145.837 & 1803.977 \\
        hydraulic & 1469.136 & 4612.781 & 34.909 & \bfseries 1.08e+04 & 923.561 & 3801.743 & 3350.071 & 3656.773 \\
        lorenz-attractor & 992.643 & 17.658 & 143.971 & 580.847 & 1650.040 & \bfseries 2191.928 & 2050.461 & 1825.448 \\
        mocap & 445.091 & 2179.353 & 148.563 & \bfseries 1.13e+05 & 1823.373 & 2525.665 & 157.968 & 1.80e+04 \\
        occupancy & 5154.791 & 9190.251 & 282.272 & 1.32e+04 & \bfseries 1.63e+04 & 7020.677 & 9255.468 & 5114.837 \\
        own-synth & 2255.508 & 74.232 & 6.455 & \bfseries 2.29e+04 & 1165.953 & 2257.976 & 1702.116 & 1026.830 \\
        thomas-attractor & \bfseries 2012.217 & nan & 8.167 & nan & 1764.227 & 1859.162 & 1352.554 & 1691.685 \\
        \bottomrule
        \end{tabular}
\end{table*}

\begin{table*}
    \footnotesize
    \centering
    \caption{Best metric 'metrics.davies-bouldin' value for each dataset and algorithm from hyperparameter search results}
    \label{tab:hp_metrics_metrics.davies-bouldin}
    \begin{tabular}{lrrrrrrrr}
        \toprule
        algorithm & birch & bocpdetector & clustream & dbstream & denstream & mini-batch-kmeans & streamkmeans & abimca \\
        dataset &  &  &  &  &  &  &  &  \\
        \midrule
        bee-waggle & 71.078 & \bfseries 1.424 & 84.739 & 5.765 & 14.012 & 19.554 & 24.979 & 12.578 \\
        cmapss & 40.052 & \bfseries 0.469 & 92.083 & 3.074 & 46.712 & 14.087 & 31.323 & 16.061 \\
        eigen-worms & 17.463 & \bfseries 0.866 & 21.523 & 4.476 & 3.846 & 2.835 & 10.923 & 3.828 \\
        hydraulic & 68.132 & \bfseries 2.947 & 84.968 & 36.736 & 152.522 & 138.098 & 45.246 & 31.007 \\
        lorenz-attractor & 38.516 & \bfseries 0.981 & 54.983 & 14.980 & 28.773 & 4.328 & 20.754 & 22.631 \\
        mocap & 35.612 & \bfseries 0.173 & 8.030 & 1.210 & 3.390 & 4.812 & 2.005 & 17.499 \\
        occupancy & 65.075 & \bfseries 5.129 & 168.704 & 6.199 & 20.439 & 15.437 & 40.378 & 23.013 \\
        own-synth & 27.162 & 8.29e+04 & 137.923 & 3.596 & 20.822 & \bfseries 2.518 & 1895.422 & 18.762 \\
        thomas-attractor & 189.853 & nan & 82.745 & nan & 5.738 & \bfseries 1.940 & 14.247 & 5.355 \\
        \bottomrule
    \end{tabular}
\end{table*}

\begin{table*}
    \footnotesize
    \centering
    \caption{Best metric 'metrics.silhouette' value for each dataset and algorithm from hyperparameter search results}
    \label{tab:hp_metrics_metrics.silhouette}
    \begin{tabular}{lrrrrrrrr}
        \toprule
        algorithm & birch & bocpdetector & clustream & dbstream & denstream & mini-batch-kmeans & streamkmeans & abimca \\
        dataset &  &  &  &  &  &  &  &  \\
        \midrule
        bee-waggle & 0.181 & -0.104 & -0.067 & 0.092 & \bfseries 0.365 & 0.192 & 0.231 & 0.318 \\
        cmapss & \bfseries 0.667 & 0.007 & 0.488 & 0.495 & 0.636 & 0.511 & 0.320 & 0.570 \\
        eigen-worms & 0.171 & 0.025 & -0.025 & 0.108 & \bfseries 0.291 & 0.247 & 0.172 & 0.272 \\
        hydraulic & 0.684 & -0.272 & -0.087 & -0.012 & 0.636 & \bfseries 0.775 & 0.769 & 0.769 \\
        lorenz-attractor & 0.380 & -0.117 & 0.031 & 0.270 & 0.349 & 0.399 & 0.387 & \bfseries 0.432 \\
        mocap & 0.233 & 0.020 & 0.054 & 0.386 & \bfseries 0.482 & 0.429 & 0.125 & 0.436 \\
        occupancy & 0.497 & -0.424 & -0.184 & \bfseries 0.774 & 0.766 & 0.647 & 0.598 & 0.484 \\
        own-synth & 0.306 & -0.233 & -0.052 & \bfseries 0.439 & 0.346 & 0.396 & 0.380 & 0.383 \\
        thomas-attractor & \bfseries 0.299 & nan & -0.076 & nan & 0.269 & 0.282 & 0.203 & 0.274 \\
        \bottomrule
    \end{tabular}
\end{table*}

\begin{table*}
    \footnotesize
    \centering
    \caption{Metric 'metrics.mt3scm' value for each dataset and algorithm from default calibration results}
    \label{tab:default_metrics_metrics.mt3scm}
    \begin{tabular}{lrrrrrrrr}
    \toprule
    algorithm & birch & bocpdetector & clustream & dbstream & denstream & mini-batch-kmeans & streamkmeans & abimca \\
    dataset &  &  &  &  &  &  &  &  \\
    \midrule
    bee-waggle & -0.009 & 0.143 & -0.021 & -0.100 & nan & -0.097 & 0.026 & \bfseries 0.282 \\
    cmapss & -0.104 & \bfseries 0.120 & -0.065 & 0.078 & -0.154 & -0.241 & 0.024 & 0.006 \\
    eigen-worms & -0.048 & 0.059 & -0.019 & -0.263 & \bfseries 0.323 & -0.074 & 0.017 & -0.276 \\
    hydraulic & -0.010 & 0.104 & -0.034 & -0.166 & 0.019 & -0.065 & \bfseries 0.272 & -0.232 \\
    lorenz-attractor & 0.037 & 0.110 & 0.003 & -0.287 & -0.138 & \bfseries 0.162 & 0.004 & -0.190 \\
    mocap & 0.025 & 0.047 & -0.071 & 0.174 & 0.096 & 0.050 & nan & \bfseries 0.225 \\
    occupancy & -0.113 & \bfseries 0.214 & -0.058 & -0.039 & -0.258 & -0.285 & 0.151 & -0.203 \\
    own-synth & -0.039 & 0.159 & -0.033 & nan & 0.028 & \bfseries 0.279 & 0.272 & -0.346 \\
    thomas-attractor & -0.036 & nan & -0.018 & nan & \bfseries 0.232 & 0.053 & -0.045 & -0.217 \\
    \bottomrule
    \end{tabular}
\end{table*}

\begin{table*}
\footnotesize
\centering
\caption{Metric 'metrics.calinski-harabasz' value for each dataset and algorithm from default calibration results}
\label{tab:default_metrics_metrics.calinski-harabasz}
\begin{tabular}{lrrrrrrrr}
\toprule
algorithm & birch & bocpdetector & clustream & dbstream & denstream & mini-batch-kmeans & streamkmeans & abimca \\
dataset &  &  &  &  &  &  &  &  \\
\midrule
bee-waggle & 39.626 & \bfseries 169.623 & 25.065 & 139.912 & nan & 113.178 & 43.804 & 5.369 \\
cmapss & 184.909 & \bfseries 3.84e+04 & 567.915 & 2635.822 & 248.424 & 1798.528 & 599.298 & 871.293 \\
eigen-worms & 353.541 & \bfseries 2498.356 & 74.512 & 327.072 & 1.306 & 876.894 & 117.145 & 55.120 \\
hydraulic & 210.616 & \bfseries 4612.781 & 112.687 & 1574.895 & 63.695 & 631.183 & 5.678 & 4.147 \\
lorenz-attractor & 234.222 & 17.658 & 48.203 & 80.547 & 320.720 & \bfseries 1060.803 & 190.984 & 18.462 \\
mocap & 64.989 & 2179.353 & 338.748 & 1954.337 & 1219.938 & 707.485 & nan & \bfseries 6561.213 \\
occupancy & 352.260 & \bfseries 9190.251 & 78.710 & 2738.856 & 879.927 & 3033.238 & 1099.447 & 171.655 \\
own-synth & 22.215 & 74.232 & 22.415 & nan & 42.333 & 1331.381 & \bfseries 2468.620 & 21.841 \\
thomas-attractor & 64.071 & nan & 37.748 & nan & 0.387 & \bfseries 1319.776 & 97.986 & 101.978 \\
\bottomrule
\end{tabular}
\end{table*}

\begin{table*}
\footnotesize
\centering
\caption{Metric 'metrics.davies-bouldin' value for each dataset and algorithm from default calibration results}
\label{tab:default_metrics_metrics.davies-bouldin}
\begin{tabular}{lrrrrrrrr}
\toprule
algorithm & birch & bocpdetector & clustream & dbstream & denstream & mini-batch-kmeans & streamkmeans & abimca \\
dataset &  &  &  &  &  &  &  &  \\
\midrule
bee-waggle & 30.023 & 1.424 & 11.749 & 4.098 & nan & 4.646 & 3.203 & \bfseries 0.593 \\
cmapss & 3.926 & \bfseries 0.469 & 16.567 & 1.017 & 3.670 & 1.483 & 1.798 & 2.143 \\
eigen-worms & 4.650 & 0.866 & 20.563 & 2.191 & \bfseries 0.859 & 1.878 & 5.871 & 1.481 \\
hydraulic & 5.295 & 2.947 & 21.356 & 5.834 & \bfseries 1.645 & 3.233 & 3.030 & 4.407 \\
lorenz-attractor & 4.899 & \bfseries 0.981 & 9.134 & 9.212 & 3.306 & 1.334 & 3.412 & 1.674 \\
mocap & 7.686 & \bfseries 0.173 & 1.686 & 0.531 & 0.866 & 1.447 & nan & 0.816 \\
occupancy & 4.536 & 5.129 & 35.854 & 2.850 & 2.901 & \bfseries 1.534 & 1.609 & 3.231 \\
own-synth & 8.585 & 8.29e+04 & 14.357 & nan & 2.781 & 1.258 & \bfseries 0.759 & 1.684 \\
thomas-attractor & 10.772 & nan & 14.992 & nan & 1.536 & 1.303 & 6.779 & \bfseries 0.874 \\
\bottomrule
\end{tabular}
\end{table*}

\begin{table*}
    \footnotesize
    \centering
    \caption{Metric 'metrics.silhouette' value for each dataset and algorithm from default calibration results}
    \label{tab:default_metrics_metrics.silhouette}
    \begin{tabular}{lrrrrrrrr}
    \toprule
    algorithm & birch & bocpdetector & clustream & dbstream & denstream & mini-batch-kmeans & streamkmeans & abimca \\
    dataset &  &  &  &  &  &  &  &  \\
    \midrule
    bee-waggle & 0.012 & -0.104 & -0.072 & -0.098 & nan & 0.015 & 0.061 & \bfseries 0.216 \\
    cmapss & 0.139 & 0.007 & -0.280 & \bfseries 0.494 & -0.168 & -0.084 & 0.319 & 0.357 \\
    eigen-worms & 0.082 & 0.025 & -0.046 & 0.018 & 0.016 & \bfseries 0.112 & 0.042 & -0.165 \\
    hydraulic & 0.128 & -0.272 & -0.113 & -0.465 & -0.207 & 0.007 & \bfseries 0.264 & -0.525 \\
    lorenz-attractor & 0.112 & -0.117 & -0.064 & -0.099 & -0.169 & \bfseries 0.211 & 0.117 & -0.416 \\
    mocap & 0.031 & 0.020 & 0.008 & 0.209 & \bfseries 0.416 & 0.231 & nan & 0.375 \\
    occupancy & 0.094 & -0.424 & -0.106 & -0.053 & -0.438 & -0.110 & \bfseries 0.126 & -0.301 \\
    own-synth & 0.012 & -0.233 & -0.051 & nan & -0.475 & 0.387 & \bfseries 0.505 & -0.505 \\
    thomas-attractor & -0.003 & nan & -0.045 & nan & -0.177 & \bfseries 0.192 & 0.019 & -0.290 \\
    \bottomrule
    \end{tabular}
\end{table*}

\begin{table*}
    \footnotesize
    \centering
    \caption{Hyperparameter random grid search upper and lower bound for each algorithm and their specific parameter options}
    \label{tab:hp_search_params}
    \begin{tabular}{lllr}
        \toprule
        &  &  & value \\
        algorithm & parameter & bound &  \\
        \midrule
        \multirow[t]{8}{*}{birch} & threshold & lower & 0.000 \\
        & branching-factor & lower & 2.000 \\
        & n-clusters & lower & 2.000 \\
        & seq-len & lower & 1.000 \\
        & threshold & upper & 1.000 \\
        & branching-factor & upper & 100.000 \\
        & n-clusters & upper & 20.000 \\
        & seq-len & upper & 30.000 \\
        \multirow[t]{4}{*}{streamkmeans} & chunk-size & lower & 1.000 \\
        & n-clusters & lower & 2.000 \\
        & chunk-size & upper & 50.000 \\
        & n-clusters & upper & 50.000 \\
        \multirow[t]{12}{*}{abimca} & learning-rate & lower & 0.000 \\
        & omega & lower & 5.000 \\
        & step-size & lower & 1.000 \\
        & seq-len & lower & 5.000 \\
        & eta & lower & 0.001 \\
        & theta-factor & lower & 1.000 \\
        & learning-rate & upper & 0.010 \\
        & omega & upper & 15.000 \\
        & step-size & upper & 3.000 \\
        & seq-len & upper & 20.000 \\
        & eta & upper & 10.000 \\
        & theta-factor & upper & 3.000 \\
        \multirow[t]{8}{*}{clustream} & time-window & lower & 1.000 \\
        & max-micro-clusters & lower & 1.000 \\
        & n-macro-clusters & lower & 1.000 \\
        & micro-cluster-r-factor & lower & 1.000 \\
        & time-window & upper & 20.000 \\
        & max-micro-clusters & upper & 10.000 \\
        & n-macro-clusters & upper & 50.000 \\
        & micro-cluster-r-factor & upper & 4.000 \\
        \multirow[t]{10}{*}{dbstream} & clustering-threshold & lower & 0.100 \\
        & fading-factor & lower & 0.001 \\
        & cleanup-interval & lower & 1.000 \\
        & minimum-weight & lower & 0.100 \\
        & intersection-factor & lower & 0.100 \\
        & clustering-threshold & upper & 5.000 \\
        & fading-factor & upper & 10.000 \\
        & cleanup-interval & upper & 10.000 \\
        & minimum-weight & upper & 5.000 \\
        & intersection-factor & upper & 1.000 \\
        \multirow[t]{8}{*}{mini-batch-kmeans} & n-clusters & lower & 1.000 \\
        & max-iter & lower & 1.000 \\
        & batch-size & lower & 128.000 \\
        & seq-len & lower & 1.000 \\
        & n-clusters & upper & 30.000 \\
        & max-iter & upper & 200.000 \\
        & batch-size & upper & 2048.000 \\
        & seq-len & upper & 100.000 \\
        \multirow[t]{12}{*}{denstream} & decaying-factor & lower & 0.000 \\
        & beta & lower & 0.000 \\
        & mu & lower & 0.000 \\
        & epsilon & lower & 0.000 \\
        & n-samples-init & lower & 1.000 \\
        & stream-speed & lower & 1.000 \\
        & decaying-factor & upper & 1.000 \\
        & beta & upper & 5.000 \\
        & mu & upper & 5.000 \\
        & epsilon & upper & 1.000 \\
        & n-samples-init & upper & 1000.000 \\
        & stream-speed & upper & 1000.000 \\
        \multirow[t]{6}{*}{bocpdetector} & lag & lower & 1.000 \\
        & changepoint-prior & lower & 0.001 \\
        & threshold & lower & 0.000 \\
        & lag & upper & 50.000 \\
        & changepoint-prior & upper & 10.000 \\
        & threshold & upper & 5.000 \\
        \bottomrule
    \end{tabular}
\end{table*}

\cleardoublepage
\printbibliography

\end{document}